\documentclass[12pt]{article}
\usepackage[utf8]{inputenc}
\usepackage[british]{babel}
\usepackage{cmap}
\usepackage{lmodern}

\usepackage{amssymb, amsmath, amsthm}
\usepackage[a4paper,top=25mm,bottom=25mm,left=25mm,right=25mm]{geometry}
\usepackage{ragged2e}

\usepackage{authblk} 
\usepackage{pifont}
\usepackage{graphicx}
\usepackage[dvipsnames,svgnames,table]{xcolor}
\usepackage[figuresright]{rotating}
\usepackage{xtab} 
\usepackage{longtable} 
\usepackage{multirow}
\usepackage{footnote}
\usepackage[stable]{footmisc}
\usepackage{chngpage} 
\usepackage{pdflscape} 
\usepackage{tocbibind} 

\usepackage{pgfplots}
\pgfplotsset{compat=1.18}
\pgfplotsset{every tick label/.append style={font=\footnotesize}}
\usepackage{setspace}

\makesavenoteenv{tabular}
\usepackage{tabularx}
\usepackage{booktabs}
\usepackage{threeparttable} 
\usepackage[referable]{threeparttablex} 
\newcolumntype{R}{>{\raggedleft\arraybackslash}X}
\newcolumntype{L}{>{\raggedright\arraybackslash}X}
\newcolumntype{C}{>{\centering\arraybackslash}X}
\newcolumntype{A}{>{\columncolor{gray!25}}C}
\newcolumntype{a}{>{\columncolor{gray!25}}c}

\newlength{\tablen}

\usepackage{dcolumn} 
\newcolumntype{.}{D{.}{.}{-1}}

\usepackage{tikz}
\usetikzlibrary{arrows, calc, matrix, patterns, positioning, trees}
\usepackage[semicolon]{natbib}
\usepackage[hyphens]{url}
\usepackage{hyperref} 
\hypersetup{
  colorlinks   = true,    		
  urlcolor     = blue,    		
  linkcolor    = blue,			
  citecolor    = ForestGreen,		  	
}
\usepackage{microtype}
\usepackage[justification=centering]{caption} 

\usepackage[labelformat=simple]{subcaption}

\DeclareCaptionLabelFormat{parenthesis}{(#2)}
\makeatletter
\renewcommand\p@subfigure{\arabic{figure}.}
\makeatother

\DeclareCaptionLabelFormat{parenthesis}{(#2)}
\makeatletter
\renewcommand\p@subtable{\arabic{table}.}
\makeatother

\usepackage{alphalph}

\def\addlegendimage{\csname pgfplots@addlegendimage\endcsname}

\usepackage{enumitem}

\setlist[itemize]{leftmargin=2.5\parindent}
\setlist[enumerate]{leftmargin=2.5\parindent}

\theoremstyle{plain}

\newtheorem{proposition}{Proposition}
\newtheorem{result}{Result}
\newtheorem{theorem}{Theorem}

\theoremstyle{definition}

\newtheorem{example}{Example}

\theoremstyle{remark}

\newtheorem{remark}{Remark}

\def\keywords{\vspace{.5em} 
{\noindent \textit{Keywords}: }}

\def\AMS{\vspace{.5em} 
{\noindent \textbf{\emph{MSC} class}: }}

\def\JEL{\vspace{.5em} 
{\noindent \textbf{\emph{JEL} classification number}: }}

\title{Random matching in balanced bipartite graphs: \\ The (un)fairness of draw mechanisms used in sports}
\author{\href{https://sites.google.com/view/laszlocsato}{L\'aszl\'o Csat\'o}\thanks{~E-mail: \emph{laszlo.csato@sztaki.hun-ren.hu}} }
\affil{Institute for Computer Science and Control (SZTAKI) \\
Hungarian Research Network (HUN-REN) \\
Laboratory on Engineering and Management Intelligence \\
Research Group of Operations Research and Decision Systems}
\affil{Corvinus University of Budapest (BCE) \\
Institute of Operations and Decision Sciences \\
Department of Operations Research and Actuarial Sciences}
\affil{Budapest, Hungary}
\date{\today}

\def\Dedication{
{\noindent
``\emph{In these circumstances, any non-random assignment of resources is likely to be asymmetric and perceived as unfair. Randomization can restore symmetry, and thus a measure of fairness.}''\footnote{~Source: \citet[p.~585]{BudishCheKojimaMilgrom2013}.}
}}

\begin{document}

\newgeometry{top=25mm,bottom=25mm,left=25mm,right=25mm}
\maketitle
\thispagestyle{empty}
\Dedication

\begin{abstract}
\noindent
The draw of some knockout tournaments requires finding a perfect matching in a balanced bipartite graph. The problem becomes challenging with draw constraints: the two draw procedures used in sports are known to be non-uniformly distributed (the feasible matchings are not equally likely), which may threaten fairness. We compare the biases of both mechanisms, each of them having two forms, for reasonable subsets of balanced bipartite graphs up to 16 nodes. An optimal mechanism exist in the draw of quarterfinals under reasonable restrictions. The UEFA Champions League Round of 16 draw is verified to apply the least distorted design among the four available options between the 2003/04 and 2023/24 seasons. However, a considerable scope remains to improve these randomisation procedures. 
\end{abstract}


\AMS{62-08, 90-10, 90B90, 91B14}

\JEL{C44, C63, Z20}

\clearpage
\restoregeometry

\section{Introduction} \label{Sec1}

The knockout competition is one of the basic tournament formats n sports \citep{DevriesereCsatoGoossens2025, ScarfYusofBilbao2009, SziklaiBiroCsato2022}. In these tournaments, two sets of teams should often be paired randomly against each other. Other constraints can be formulated by the organiser to inrease attractiveness and ensure diversity. Perhaps the most famous example is the UEFA Champions League Round of 16 draw between the 2003/04 and 2023/24 seasons, where the eight group winners and the eight group runners-up have been matched against each other such that the assignment has not been allowed to contain any pair of teams coming from the same group or national association \citep{KlossnerBecker2013, BoczonWilson2023}.

This leads to a well-known mathematical problem: finding a perfect matching in a balanced bipartite graph \citep{HopcroftKarp1973, TanimotoItaiRodeh1978}. In order to guarantee the fairness of the draw, any team needs to play against any of its possible opponent with the same probability as if a perfect matching would be chosen randomly. Otherwise, a team will get stronger opponents in expected terms and might feel cheated.

In theory, the required property can be easily achieved by either choosing uniformly from the set of all valid assignments (if the number of nodes is relatively small), or using a uniform rejection sampler (that produces a random perfect matching of the associated complete bipartite graph, which is repeated until the matching satisfies all constraints).
However, these fair algorithms suffer from two weaknesses:
(1) the drama or entertainment value of the draw would be lost; and
(2) all stakeholders should ``trust'' the computer to sample correctly.
Even though the first issue can be solved via a Metropolis or swap algorithm \citep{KlossnerBecker2013, RobertsRosenthal2024}, transparency remains a challenging requirement. It would be impossible to detect fraud and prevent conspiracy theories if the computer generates a perfect matching that is said random but certainly favours some teams at the expense of others. 

Therefore, the draw mechanism should also be perceived and understood as fair by all participants, which necessitates to verify all calculations at least \emph{ex-post} \citep{BoczonWilson2023}. For instance, a mistake was recognised in the 2021/22 UEFA Champions League Round of 16 draw, and the process was repeated three hours later \citep{Guyon2021d}.

According to our knowledge, two field-proven draw procedures exist in the realm of sports that can be used to choose a random perfect matching in a balanced bipartite graph.
The first, applied in European club football competitions, is called here the \emph{Drop mechanism}. The second, applied in the group draw of various tournaments, including the FIBA Basketball World Cup and the FIFA World Cup, is called the \emph{Skip mechanism}.
In addition, both the Drop and Skip mechanisms have two forms depending on the side of the bipartite graph where they start, similar to the well-known Gale--Shapley algorithm.

Naturally, there is no ``free lunch'' and all these transparent draw procedures are non-uniform, i.e., the feasible outcomes are not equally likely. The previous literature has already analysed the distortions of the Drop \citep{Kiesl2013, KlossnerBecker2013, BoczonWilson2023} and the Skip \citep{Csato2025c, RobertsRosenthal2024} mechanism in different settings. However, they have \emph{never} been compared, which will be presented in the current paper. Since analytical calculations are cumbersome and should be done separately for each graph, our study is mainly limited to simulation results, analogous to the papers listed above.

Our main contributions can be summarised as follows:
\begin{itemize}
\item
The smallest bipartite graph is found for which the two draw mechanisms are unfair (Proposition~\ref{Prop1});
\item
The Skip mechanism is proved to dominate the Drop mechanism in a draw of quarterfinals under reasonable restrictions (Theorem~\ref{Theo1});
\item
The mechanism used for the UEFA Champions League Round of 16 draw is documented to be the least distorted among the four known transparent and field-proven ptions (Results~\ref{Result3} and \ref{Result4});
\item
The four draw mechanisms are demonstrated to distort the probabilities in the same direction and roughly with the same magnitude (Result~\ref{Result7}).
\end{itemize}
Consequently, if the draw constraints are analogous to the restrictions applied in the UEFA Champions League, the Skip mechanism is the optimal choice for the draw of the quarterfinals but the Drop mechanism stochastically dominates it for the draw of the Round 16. Result~\ref{Result2} offers a potential explanation as increasing the number of teams without additional draw restrictions decreases the bias of the Drop mechanism more rapidly than the bias of the Skip mechanism.

We also substantially refine existing results on the UEFA Champions League Round of 16 draw. \citet{BoczonWilson2023} and \citet{KlossnerBecker2013} have not identified the advantage of the Drop mechanism over the Skip mechanism since not considering the latter.
Both \citet[Footnote~33]{BoczonWilson2023} and \citet[Footnote~19]{KlossnerBecker2013} suggest that the difference between the Standard and Reversed Drop mechanisms is slight and uninteresting, which can be debated (Results~\ref{Result1} and \ref{Result6}). 

What are the lessons for practitioners, such as tournament organisers?
Usually, some different transparent randomisation procedures exist for a given allocation or assignment problem. It is advised to consider all these mechanisms and choose the most appropriate lottery for the given setting after analysing their performance with respect to various theoretical properties such as fairness.

This would be important even if the effect sizes found seem to be small at first glance. First, the biases can imply quite substantial financial differences: in the UEFA Champions League Round of 16 draw, the expected revenue of some teams is usually changed by more than 10 thousand euros in almost every season because of the distortions \citep{KlossnerBecker2013}. Second, changing the draw procedure to a better field-proven alternative has essentially no price, thus, it is worth implementing essentially independently of the extent of improvement.

\section{Related literature} \label{Sec2}

Our paper is connected to two distinct lines of literature: the theory of random allocation mechanisms and fairness in tournament design.

The \emph{assignment problem}, widely discussed in the field of market design, aims to allocate indivisible objects among agents such that any agent receives exactly one object, and monetary transfers are prohibited. Examples include assigning of jobs to workers or time slots to users of a common good. Using a lottery is a straightforward trick for guaranteeing fairness: the random priority mechanism draws a random ordering of the agents from the uniform distribution and lets them successively choose an object \citep{AbdulkadirougluSonmez1998}. However, while uniform draw is traditionally used in sports as will turn out in Section~\ref{Sec3}, the policy of choosing the opponents is rarely applied \citep{Guyon2022a, HallLiu2024, LunanderKarlsson2023}.

\citet{HyllandZeckhauser1979} and \citet{BogomolnaiaMoulin2001} have developed an alternative solution to random assignment by directly identifying an expected assignment matrix that contains the probabilities with which the agents should get the objects.
\citet{BudishCheKojimaMilgrom2013} expand this theory to a much wider class of matching and assignment environments to accommodate multi-unit allocations and various real-world constraints, such as different quotas.
\citet{BoczonWilson2023} verify the power of this market design tool in the normative assessment of the UEFA Champions League Round of 16 draw.

The second approach, finding a procedure to reproduce an expected assignment matrix is natural in our setting since the ideal probabilities under uniform distribution can be easily calculated by a uniform rejection sampler \citep{RobertsRosenthal2024}. Indeed, \citet{BoczonWilson2023} conclude that the design of the UEFA Champions League Round of 16 draw is near-optimal and some reasonable variants of it are unable to reduce unfairness. However, they do not consider the Skip mechanism at all, which will be done in the following. Furthermore, the two variants of the Drop mechanism have the same level of fairness according to \citet[Footnote~33]{BoczonWilson2023} but this is a simplification according to Sections~\ref{Sec55} and \ref{Sec56}.

The draw of sports tournaments has been analysed in operational research and statistics, too. Before \citet[Proposition~2]{BoczonWilson2023}, the non-uniformity of the Drop mechanism has already been demonstrated by \citet{Guyon2014a, Kiesl2013, KlossnerBecker2013, WallaceHaigh2013}.
Thus, \citet{RobertsRosenthal2024} have proposed various algorithms that guarantee uniform distribution---but they are more difficult to implement and understand than the Drop and Skip mechanisms.
Analogously, \citet{LalienaLopez2025} introduce three different draw procedures based on pots and balls in order to extract one feasible solution from a given list of balanced assignments.

Since draw constraints can be used in tournaments to achieve certain strategic goals such as minimising the threat of unfair behaviour \citep{Csato2022a}, the number and complexity of these conditions are expected to increase in the future. Section~\ref{Sec32} will bring some evidence for this development, which probably increases the role of the draw procedure. 

Random matching in a balanced bipartite graph makes sense only if at least one valid assignment exists. That can be checked by Hall's marriage theorem as shown by \citet{Kiesl2013} and \citet{WallaceHaigh2013} for the UEFA Champions League Round of 16 draw.

Finally, in contrast to the assignment problem in market design, the preferences of teams are ignored in sports competitions (and in our paper, too) since the organiser does not want to maximise welfare but approximate uniform distribution, guaranteeing the equal probability of all outcomes. An alternative interpretation can be that the teams have the same preference ordering on the set of their possible opponents because, for example, the strength of any team is given by a real number, which is common knowledge.

\section{Randomisation procedures used in sports} \label{Sec3}

It is a common problem in sports to divide a set of contestants into groups. The allocation is often subject to some rules, such as the existence of seeding pots to guarantee balancedness \citep{LalienaLopez2019, LaprePalazzolo2023}, and different criteria that ensure geographical diversity for increasing attractiveness \citep{Guyon2015a}. The seeding pots themselves do not imply any challenge since balls representing the teams can be drawn sequentially from urns associated with the seeding pots, which ensures uniform distribution. However, the presence of other constraints makes some assignments invalid, and the draw procedure should produce an admissible matching. According to our knowledge, two draw mechanisms are used in practice to that end. They are presented in Sections~\ref{Sec31} and \ref{Sec32}, respectively.

\subsection{The UEFA Champions League Round of 16 draw and the Drop mechanism} \label{Sec31}

The UEFA Champions League is the most prestigious European football club competition. Teams can qualify primarily based on the results of the domestic leagues in the previous year. Since the 1997/98 season, multiple entrants are allowed from the top national leagues.

The Champions League has been organised in the same format between the 2003/04 and 2023/24 seasons: a group stage played in eight home-away round-robin groups, where the top two teams qualify for the knockout stage.
In the Round of 16, the eight group winners and the eight runners-up are divided into eight mutually disjoint pairs subject to the following restrictions:
\begin{itemize}
\item
\emph{Bipartite constraint}: a group winner should play against a runner-up;
\item
\emph{Group constraint}: teams from the same group are not allowed to be paired;
\item
\emph{Association constraint}: clubs from the same country cannot be matched.
\end{itemize}
The first restriction reduces the number of valid assignments to $8! = 40{,}320$ as every outcome can be represented by a permutation of the eight groups.
The second restriction means that only a derangement (a permutation without a fixed point, where no element appears in its original position) is allowed, which decreases the number of feasible solutions to $!8 = \lfloor 8! / e + 1/2 \rfloor = 14{,}833$ \citep{KlossnerBecker2013} (the number of derangements of an $n$-element set is denoted by $!n$).
Finally, the impact of the association constraint depends on the identity of the teams, and no simple combinatorial formula exists to determine the number of possible draw outcomes. For instance, there have been $10{,}595$ feasible assignments in the 2023/24 season, but only $3{,}876$ in the 2022/2023 season.

UEFA has used the following mechanism in the Round of 16 draw:
\begin{itemize}
\item
Eight balls containing the names of the eight runners-up are placed in a bowl;
\item
A ball is drawn from the bowl, and the team drawn plays at home in match 1;
\item
The computer shows which group winners are eligible to play as the visiting team in match 1;
\item
Balls representing these teams are placed in another bowl;
\item
A ball is drawn from the second bowl to complete the pairing for match 1;
\item
The above procedure is repeated for the remaining matches.
\end{itemize}
The computer may indicate that only one group winner is allowed to play as the visiting team when this team is automatically assigned to the match.

In the following, this procedure will be called the \emph{Drop mechanism}.
It has two forms, depending on the side of the bipartite graph where it is started: UEFA draws the runner-up first, but the group winner can also be drawn first.

The Drop mechanism is more complicated than it seems at first glance because it should be checked not only whether draw conditions apply for the runner-up chosen at the moment, but also whether draw conditions are anticipated to apply for the teams still to be drawn. Let us see an illustration.

\begin{table}[t!]
  \centering
  \caption{Teams playing in the 2012/13 UEFA Champions League Round of 16}
  \label{Table1}
    \rowcolors{1}{gray!20}{}
    \begin{tabularx}{0.9\textwidth}{cLlLl} \toprule \hiderowcolors
    \multirow{2}[0]{*}{Group} & \multicolumn{2}{c}{Runner-up} & \multicolumn{2}{c}{Group winner} \\
          & Club  & Country & Club  & Country \\ \bottomrule \showrowcolors
    A     & Porto & Portugal & Paris Saint-Germain & France \\
    B     & Arsenal & England & Schalke 04 & Germany \\
    C     & Milan & Italy & M\'alaga & Spain \\
    D     & Real Madrid & Spain & Borussia Dortmund & Germany \\
    E     & Shakhtar Donetsk & Ukraine & Juventus & Italy \\
    F     & Valencia & Spain & Bayern M\"unchen & Germany \\
    G     & Celtic Glasgow & Scotland & Barcelona & Spain \\
    H     & Galatasaray & Turkey & Manchester United & England \\ \bottomrule
    \end{tabularx}
\end{table}

\begin{example} \label{Examp1} \citep{KlossnerBecker2013}
The participants of the 2012/13 UEFA Champions League Round of 16 are shown in Table~\ref{Table1}.
The draw happened as follows:
\begin{enumerate}
\item
The runner-up Galatasaray was drawn first. Its eligible opponents were all teams except for Manchester United owing to the group constraint. Out of the seven group winners, Schalke 04 was drawn.
\item
The runner-up Celtic Glasgow was drawn second. Its possible opponents were all teams except for Schalke 04 (already drawn), and Barcelona (group constraint). Out of the six group winners, Juventus was drawn.
\item
The runner-up Arsenal was drawn third. Its admissible opponents were all teams except for Schalke 04, Juventus (already drawn), and Manchester United (association constraint). The group constraint was ineffective. Out of the five group winners, Bayern M\"unchen was drawn.
\item
The runner-up Shakhtar Donetsk was drawn fourth. Its eligible opponents were all teams except for Schalke 04, Juventus, and Bayern M\"unchen (already drawn). Out of the five remaining group winners, Borussia Dortmund was drawn.
\item
The runner-up Milan was drawn fifth. The computer indicated that it should be paired with Barcelona, hence, no draw was carried out.
\end{enumerate}
Does this indicate a flaw? Naturally not. Four group winners (Paris Saint-Germain, M\'alaga, Barcelona, Manchester United) remained to be drawn. M\'alaga was prohibited by the group constraint. If Milan had played against the French or English team, then three pairings would have been left with four Spanish clubs (Real Madrid, Valencia, M\'alaga, Barcelona), and the association constraint would have been violated.

However, two (Real Madrid, Valencia) out of the three remaining runners-up still could have faced either Paris Saint-Germain or Manchester United. Therefore, the draw was not finished in the fifth round after Milan was drawn.
\end{example}

The same algorithm is used in other UEFA club competitions, too. The UEFA Cup Round of 32 draw was organised with both the association and the group constraints from 2004/05 until its last 2008/09 season \citep{UEFA2004}. The competition was renamed to UEFA Europa League and retained both restrictions in the Round of 32 draw until the 2020/21 season \citep{UEFA2009b, Csato2022d}. Between the 2021/22 and 2023/24 seasons, both the UEFA Europa League and the UEFA Europa Conference League contained knockout round play-offs contested by 16 teams as well as the Round of 16, where the association constraint applied but the group constraint was not considered \citep{UEFA2021e, UEFA2021d}.

A live probability calculator for the UEFA club competitions in the 2023/24 season, based on the Drop mechanism, is available at \url{https://julienguyon.github.io/UEFA-draws/}.
\citet{GuyonMeunier2023} summarise the mathematical background of the calculator.

\subsection{The traditional algorithm for restricted group draw: the Skip mechanism} \label{Sec32}

The FIFA World Cup is intended to contain geographically diverse groups since at least 1990. However, the draw of the 1990 \citep{Jones1990}, 2006 \citep{RathgeberRathgeber2007}, and 2014 \citep{Guyon2015a} tournaments were seriously unfair. Hence, the French mathematician \emph{Julien Guyon} has proposed using the Drop mechanism for the FIFA World Cup draw \citep{Guyon2014a} (this recommendation is missing from the published version of \citet{Guyon2015a}). FIFA has heard the message and adopted a similar, credible and transparent draw procedure \citep{Guyon2018d}: the team drawn is assigned to the first available slot in alphabetical order as indicated by the computer, meaning that a free slot may be ``skipped''. Let us see an illustration.

\begin{example}
Take the 2012/13 UEFA Champions League Round of 16 draw with the teams presented in Table~\ref{Table1}. Assume that the runners-up are drawn first in the order Galatasaray, Celtic Glasgow, Arsenal, Shakhtar Donetsk, Milan, Real Madrid, Porto, and Valencia. After the urn of the runners-up is emptied, the draw continues with the urn of the group winners.
\begin{enumerate}
\item
Schalke 04 is drawn first and matched with Galatasaray.
\item
Juventus is drawn second and assigned to Celtic Glasgow.
\item
Manchester United is drawn third and matched with Shakhtar Donetsk (the association constraint applies to Arsenal).
\item
Bayern M\"unchen is drawn fourth and assigned to Arsenal, which has had no opponent before.
\item
Borussia Dortmund is drawn fifth and matched with Porto.
\end{enumerate}
Why does Borussia Dortmund skip both Milan and Real Madrid? The situation is analogous to Example~\ref{Examp1}. Three group winners (Paris Saint-Germain, M\'alaga, Barcelona) remain to be drawn. If Milan plays against Borussia Dortmund, then three pairings will be left with four Spanish teams, and the association constraint will be violated. Furthermore, Real Madrid is not allowed to play against Borussia Dortmund due to the group constraint.
\end{example}

This procedure will be called the \emph{Skip mechanism}.

It is used more extensively than the Drop mechanism, including the following competitions:
\begin{itemize}
\item
FIBA Basketball World Cup: 2019 \citep{FIBA2019}, 2023 \citep{FIBA2023};
\item
FIFA World Cup: 2018 \citep{FIFA2017c}, 2022 \citep{FIFA2022a};
\item
European qualifiers for the FIFA World Cup: 2018 \citep{UEFA2015f}, 2022 \citep{UEFA2020c}, 2026 \citep{FIFA2024};
\item
UEFA Euro qualifying: 2016 \citep{UEFA2014b}, 2020 \citep{UEFA2018d}, 2024 \citep{UEFA2022e};
\item
UEFA Nations League: 2018/19 \citep{UEFA2018b}, 2020/21 \citep{UEFA2020d}, 2022/23 \citep{UEFA2021i}, 2024/25 \citep{UEFA2024a}.
\end{itemize}
Interestingly, the set of constraints has somewhat evolved over the years. For example, the 2016 UEFA Euro qualifying draw and the draw of the European qualifiers for the 2018 FIFA World Cup contained only some prohibited clashes and required five teams to be drawn into larger groups, while the 2020 UEFA Euro qualifying draw applied restrictions due to host nations, prohibited clashes, winter venue, and excessive travel.

Naturally, the Skip mechanism has as many variants as the possible orders of the pots \citep{Csato2025c}. It usually starts with the pot containing the strongest teams, although the European qualifiers for the 2018 FIFA World Cup and the UEFA Nations League until 2022/23 have followed a reversed order. For bipartite graphs, the Skip mechanism has two different forms.

\section{Methodology} \label{Sec4}

In the following, the methodology of our study is detailed.
Section~\ref{Sec41} presents the basic mathematical notions connected to the setting. Section~\ref{Sec42} overviews the graphs on which the draw mechanisms described in Section~\ref{Sec43} are investigated. Last but not least, Section~\ref{Sec44} discusses the comparison metrics.

\subsection{Preliminaries} \label{Sec41}

The nodes of a bipartite graph can be divided into two disjoint and independent sets $U$ and $V$ such that every edge is between a node in $U$ and a node in $V$. A bipartite graph is called \emph{balanced} if $|U| = |V| = n$. A \emph{perfect matching} is an independent edge set of size $n$.

For every bipartite graph $G$, the set of prohibited pairs, that is, the complement of graph $G$ will be considered. There are two kinds of constraints: \emph{pair constraints} and \emph{type constraints}.
Each node in $U$ has a pair in $V$ and vice versa, and it might not be allowed for the perfect matching to contain these $n$ edges called pair constraints.
Furthermore, every node has a type, which is different from the type of its pair. Type constraints imply that the perfect matching cannot contain any edge between two nodes of the same type.

In the complement of a graph $G$, the degree sequence of the set $U$ ($V$) lists the degrees of nodes in $U$ ($V$) in a non-increasing order.
Without loss of generality, the degree sequence of the set $U$ is assumed \emph{not} to be lexicographically smaller than the degree sequence of the set $V$ in the complement of any balanced bipartite graph $G$.
In all figures, nodes of $U$ will be on the left-hand side, while nodes of $V$ on the right-hand side. However, set $U$ contains the groups winners and set $V$ consists of the runners-up for the historical UEFA Champions League seasons.

\subsection{The choice of balanced bipartite graphs} \label{Sec42}

Three kinds of balanced bipartite graphs are investigated.
The first set consists of all graphs for $n=4$, both with and without pair constraints. The impact of adding one or two pairs of nodes with no additional type constraints will also be evaluated.
The second set is provided by the 21 historical UEFA Champions League seasons between 2003/04 and 2023/24.
The third set contains the graph corresponding to the 2017/18 Champions League Round of 16 (where set $U$ contains the runners-up since their degree sequence is lexicographically higher), and seven graphs of the same pattern.

\input{Figure1_n=4_all_cases}

For $n = 4$, Figure~\ref{Fig1} presents the 31 possible graphs $G_{1}$--$G_{31}$ with pair constraints if the type of each node and its pair is different.
The last four graphs ($G_{28}$--$G_{31}$) have valid assignments only if $n \geq 5$.
These graphs may represent the draw of quarterfinals in a sports tournament, where four group winners and four runners-up should be matched such that teams from the same group cannot play against each other and no group contains more than one team of each type.

\input{Figure2_n=4_all_cases_noG}

For $n = 4$, Figure~\ref{Fig2} presents all the 20 possible graphs $H_{1}$--$H_{31}$ without pair constraints. Their labelling follows the graphs $G_{1}$--$G_{31}$, for instance, $H_{12-14}$ can be obtained from either $G_{12}$, or $G_{13}$, or $G_{14}$ by removing the pair constraints. 
These graphs may represent the draw of quarterfinals in a sports tournament, where four group winners and four runners-up should be matched such that teams from the same group are allowed to play against each other but no group contained more than one team from each type.

\input{Figure3_highly_unfair_graphs}

Finally, Figure~\ref{Fig3} shows eight graphs, all representing possible outcomes in the UEFA Champions League Round of 16. From at most two national associations, five clubs can play in the group stage: five German (Spanish) teams have participated in the 2022/23 (2023/24) season.

\subsection{Draw mechanisms} \label{Sec43}

The \emph{Uniform mechanism} is uniformly distributed over the set of perfect matchings, it chooses one matching randomly.

The \emph{Standard Drop mechanism} is the Drop mechanism starting from the set $V$ (nodes on the right-hand side), while the \emph{Reversed Drop mechanism} is the Drop mechanism starting from the set $U$ (nodes on the left-hand side). These names follow the convention of UEFA that starts the Drop mechanism on the side of the runners-up.

The \emph{Standard Skip mechanism} is the Skip mechanism starting from the set $U$ (nodes on the left-hand side), while the \emph{Reversed Skip mechanism} is the Drop mechanism starting from the set $V$ (nodes on the right-hand side).

\subsection{Quantifying unfairness} \label{Sec44}

The (un)fairness of the draw mechanisms will be evaluated by focusing on the probability of matching two nodes. Let $p_{ij}$ be the probability that nodes $i$ and $j$ are matched under the Uniform mechanism, and $p_{ij}^M$ be the probability of this event if mechanism $M$ is used.
Two reasonable fairness distortion measures for mechanism $M$ are:

\begin{equation} \label{eq_AFD}
\mathit{AFD}(M) = 100 \cdot \frac{\sum_{i,j} \left| p_{ij} - p_{ij}^M \right|}{\sum_{i,j} \# \left\{ p_{ij} > 0 \right\}},
\end{equation}

\begin{equation} \label{eq_MFD}
\mathit{MFD}(M) = 100 \cdot \max_{i,j} \left| p_{ij} - p_{ij}^M \right|,
\end{equation}
where $\sum_{i,j} \# \left\{ p_{ij} > 0 \right\}$ is the number of pairs with a positive probability.

$\mathit{AFD}$ is called \emph{average fairness distortion} and $\mathit{MFD}$ is called \emph{maximal fairness distortion}. Obviously, $\mathit{AFD}$ allows the draw mechanism to compensate for the increased unfairness regarding a particular pair of nodes with a reduction in other pairs, while this is not possible under $\mathit{MFD}$. Furthermore, the value of $\mathit{MFD}$ has a more straightforward interpretation: how much is the probability of a potential pair changed by mechanism $M$ compared to the uniform draw in the \emph{worst case}.
The multiplier 100 in formulas~\eqref{eq_AFD} and \eqref{eq_MFD} means that both fairness distortions can be interpreted as a percentage.
Both metrics will be illustrated in Example~\ref{Examp3}.

The maximum of $\sum_{i,j} \# \left\{ p_{ij} > 0 \right\}$ is $n(n-1)$, which is reached if the type constraint is ineffective. However, this has never happened in the UEFA Champions League, where the denominator varies between 43 (2019/20) and 54 (2023/24).

Note that not only the Uniform mechanism may be undistorted. However, if a draw procedure implies that each team has the same probability to play against any of its possible opponents as under the Uniform mechanism, no team can effectively argue against this draw procedure.

Two draw procedures $M_1$ and $M_2$ are said to \emph{coincide} for graph $G$ if $p_{ij}^{M_1} = p_{ij}^{M_2}$ for every $i \in U$ and $j \in V$.

For $n \leq 6$ (Sections~\ref{Sec51}--\ref{Sec53}), we compute the exact probabilities via complete enumeration.
If $n=8$ (Sections~\ref{Sec55} and \ref{Sec56}), the draw mechanisms are simulated $N = 10^8$ times, which leads to 95 percent confidence intervals smaller than $\pm 0.00015$ \citep[Proposition~4]{BoczonWilson2023}. This upper bound takes into account that a conservative estimate for the assignment probability is 0.5 rather than 0.25 since the probability of Barcelona versus Chelsea is 43.75\% in the 2017/18 UEFA Champions League Round of 16.

\citet{BoczonWilson2023} use a more complex quantification of fairness by computing the average absolute difference in the match likelihoods across all valid pairwise comparisons. Consequently, their distortion metric equals zero only in the absence of draw constraints, and tends to increase with the number of restricted pairs. On the other hand, both $\mathit{AFD}$ and $\mathit{MFD}$ are zero for the Uniform mechanism.

\section{Results} \label{Sec5}

This section presents our findings.
Section~\ref{Sec51} identifies the smallest balanced bipartite graph for which the Drop and Skip mechanisms are unfair.
Section~\ref{Sec52} shows the distortions for the graphs displayed in Figures~\ref{Fig1} and \ref{Fig2}.
Section~\ref{Sec53} analyses the same graphs after adding one or two pairs of nodes without further type constraints.
Based on these calculations, Section~\ref{Sec54} provides sufficient conditions for the fairness of these draw procedures.
Section~\ref{Sec55} examines the graphs provided by the 21 historical UEFA Champions League seasons, while Section~\ref{Sec56} studies the graphs shown in Figure~\ref{Fig3}.

\subsection{The unfairness of draw mechanisms: a minimal example} \label{Sec51}

The first statement uncovers the simplest case when the draw mechanisms used in sports are distributed non-uniformly. The calculation is not especially interesting, but it clearly reveals why the derivation of general analytical results is close to impossible.

\begin{proposition} \label{Prop1}
Assume that $n \leq 3$.
The Standard Drop, Reversed Drop, Standard Skip, and Reversed Skip mechanisms coincide and they are \emph{unfair} for only one particular graph.
\end{proposition}

\begin{figure}[t!]
\centering

\begin{tikzpicture}[scale=0.8, auto=center]
\tikzstyle{every node}=[draw,shape=circle];
  \node[minimum size=1cm] (n1) at (0,4) {a};
  \node[minimum size=1cm] (n2) at (0,2) {b};
  \node[minimum size=1cm] (n3) at (0,0) {c};
  \node[minimum size=1cm] (n4) at (4,4) {A};
  \node[minimum size=1cm] (n5) at (4,2) {B};
  \node[minimum size=1cm] (n6) at (4,0) {C};

  \foreach \from/\to in {n1/n4,n2/n5}
    \draw (\from) -- (\to);
\end{tikzpicture}

\captionsetup{justification=centerfirst}
\caption{The only balanced bipartite graph up to six nodes where unfairness emerges \\ \vspace{0.2cm}
\footnotesize{\emph{Note}: Solid lines indicate the type constraints.}}
\label{Fig4}
\end{figure}
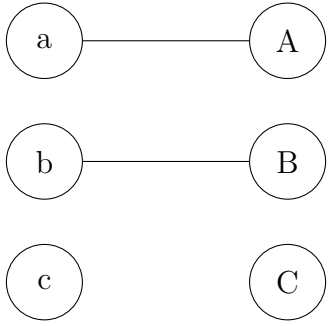


\begin{proof}
All bipartite graphs with $n \leq 3$ have been checked via complete enumeration. The unique graph for which unfairness emerges is shown in Figure~\ref{Fig4}.
In the case of all draw mechanisms, we compute the probability of nodes $a$ and $B$ being matched.

\emph{Uniform mechanism}: There are $3! = 6$ assignments in the absence of the constraints. Both the first (nodes $a$ and $A$) and the second (nodes $b$ and $B$) type constraints exclude $2! = 2$ assignments, respectively. However, $1! = 1$ among them is counted twice, thus, the number of valid assignments equals $3$. Nodes $a$ and $B$ are matched in $2! = 2$ assignments, the corresponding probability is $2/3$.

\emph{Drop mechanism}:
Five cases will be distinguished with respect to the draw order of the nodes in $U$:
\begin{itemize}
\item
Node $a$ is drawn first (this event has a chance of $1/3$):
Node $a$ has two possible pairs ($B$ and $C$), $B$ is chosen with a probability of $1/2$.
\item
Node $a$ is drawn second and node $b$ is drawn first (this event has a chance of $1/6$):
Node $b$ has two possible pairs ($A$ and $C$). \\
If $b$ is assigned to $A$, node $a$ has two possible pairs ($B$ and $C$) among the remaining nodes, $B$ is chosen with a probability of $1/2$. \\
If $b$ is assigned to $C$, node $a$ has only one possible pair ($B$) among the remaining nodes.
\item
Node $a$ is drawn second and node $b$ is drawn third (this event has a chance of $1/6$):
Node $c$ has three possible pairs. \\
If $c$ is assigned to $A$, node $a$ has two possible pairs ($B$ and $C$) among the remaining two nodes. However, $b$ cannot be assigned to $B$, hence, $B$ is chosen with a probability of $1$. \\
If $c$ is assigned to $B$, node $a$ cannot be matched with $B$. \\
If $c$ is assigned to $C$, node $a$ has only one possible pair ($B$) among the remaining two nodes.
\item
Node $a$ is drawn third and node $b$ is drawn first (this event has a chance of $1/6$):
Node $b$ has two possible pairs ($A$ and $C$). \\
If $b$ is assigned to $A$, node $c$ has two possible pairs ($B$ and $C$) among the remaining two nodes. Thus, the probability that $a$ is paired with $B$ equals $1/2$. \\
If $b$ is assigned to $C$, node $c$ has two possible pairs ($A$ and $B$) among the remaining two nodes. However, $a$ cannot be assigned to $A$, hence, the probability that $a$ is paired with $B$ equals $1$.
\item
Node $a$ is drawn third and node $b$ is drawn second (this event has a chance of $1/6$):
Node $c$ has three possible pairs. \\
If $c$ is assigned to $A$, node $b$ has only one possible pair ($C$) among the remaining two nodes, and node $a$ will certainly be matched with $B$. \\
If $c$ is assigned to $B$, node $a$ cannot be matched with $B$. \\
If $c$ is assigned to $C$, node $b$ has only one possible pair ($A$) among the remaining two nodes, and node $a$ will certainly be matched with $B$.
\end{itemize}
Consequently, the probability that node $a$ is assigned to node $B$ equals
\begin{align*}
\frac{1}{3} \cdot \frac{1}{2} + \frac{1}{6} \cdot \left( \frac{1}{2} \cdot \frac{1}{2} + \frac{1}{2} \cdot 1 \right) + \frac{1}{6} \cdot \left( \frac{1}{3} \cdot 1 + \frac{1}{3} \cdot 0 + \frac{1}{3} \cdot 1 \right) + \frac{1}{6} \cdot \left( \frac{1}{2} \cdot \frac{1}{2} + \frac{1}{2} \cdot 1 \right) + \\
+ \frac{1}{6} \cdot \left( \frac{1}{3} \cdot 1 + \frac{1}{3} \cdot 0 + \frac{1}{3} \cdot 1 \right) = \frac{1}{6} + \frac{1}{8} + \frac{1}{9} + \frac{1}{8} + \frac{1}{9} = \frac{23}{36}.
\end{align*}

\emph{Skip mechanism}:
Six cases will be distinguished with respect to the draw order of the nodes in $U$, each of them having a chance of $1/6$ to occur:
\begin{itemize}
\item
Node $a$ is drawn first and node $b$ is drawn second:
Node $B$ will be assigned to node $a$ if
(a) node $B$ is drawn first from $V$; or
(b) node $A$ is drawn first and node $B$ is drawn second from $V$ (because node $A$ ``skips'' node $a$). \\
This has a probability of $1/2$ because the nodes in $V$ have $3! = 6$ permutations.
\item
Node $a$ is drawn first and node $b$ is drawn third:
Node $B$ will be assigned to node $a$ if
(a) node $B$ is drawn first from $V$; or
(b) node $A$ is drawn first (when node $A$ ``skips'' node $a$). \\
This has a probability of $2/3$.
\item
Node $a$ is drawn second and node $b$ is drawn first:
Node $B$ will be assigned to node $a$ if
(a) node $B$ is drawn first from $V$ (because $B$ cannot matched with $b$); or
(b) node $B$ is drawn second from $V$; or
(c) node $C$ is drawn first and node $B$ is drawn third from $V$ (when node $A$ ``skips'' node $a$ and is assigned to node $c$). \\
This has a probability of $5/6$ since only the permutation $(A,C,B)$ is not appropriate.
\item
Node $a$ is drawn second and node $b$ is drawn third:
Node $B$ will be assigned to node $a$ if
(a) node $B$ is drawn second from $V$; or
(b) node $B$ is drawn third from $V$ (when the node drawn second from $V$ ``skips'' node $a$). \\
This has a probability of $2/3$.
\item
Node $a$ is drawn third and node $b$ is drawn first:
Node $B$ will be assigned to node $a$ if
(a) node $B$ is drawn second and node $A$ is drawn third from $V$ (when node $B$ ``skips'' node $c$); or
(b) node $B$ is drawn third from $V$. \\
This has a probability of $1/2$.
\item
Node $a$ is drawn third and node $b$ is drawn second:
Node $B$ will be assigned to node $a$ if
(a) node $B$ is drawn second from $V$ (when node $B$ ``skips'' node $b$); or
(b) node $B$ is drawn third from $V$. \\
This has a probability of $2/3$.
\end{itemize}
Consequently, the probability that node $a$ is assigned to node $B$ equals
\begin{align*}
\frac{1}{6} \cdot \left( \frac{1}{2} + \frac{2}{3} + \frac{5}{6} + \frac{2}{3} + \frac{1}{2} + \frac{2}{3} \right) = \frac{1}{6} \cdot \frac{23}{6} = \frac{23}{36}.
\end{align*}
The Standard and Reversed forms of the Drop and Skip mechanisms coincide since the sets $U$ and $V$ are exchangeable in the bipartite graph shown in Figure~\ref{Fig4}.
\end{proof}

\begin{table}[t!]
  \centering
  \caption{Assignment matrices for the balanced bipartite graph with six nodes in Figure~\ref{Fig4}}
  \label{Table2}
  
\begin{subtable}{\textwidth}
  \centering
  \caption{Uniform mechanism: Probabilities for all pairs of nodes}
  \label{Table2a}
    \rowcolors{1}{}{gray!20}
    \begin{tabularx}{0.6\textwidth}{l CCC} \toprule \hiderowcolors
          & A     & B     & C \\ \bottomrule \showrowcolors
    a     & 0  	  & 2/3   &  1/3 \\
    b     & 2/3   &  0    &  1/3 \\
    c     &  1/3  &  1/3  &  1/3 \\ \toprule
    \end{tabularx}
\end{subtable}

\vspace{0.25cm}
\begin{subtable}{\textwidth}
  \centering
  \caption{Drop and Skip mechanisms: Probabilities for all pairs of nodes}
  \label{Table2b}
    \rowcolors{1}{}{gray!20}
    \begin{tabularx}{0.6\textwidth}{l CCC} \toprule \hiderowcolors
          & A      & B      & C \\ \bottomrule \showrowcolors
    a     & 0      & 23/36  &  13/36 \\
    b     & 23/36  &  23/36 &  13/36 \\
    c     &  13/36 &  13/36 &   5/18 \\ \toprule
    \end{tabularx}
\end{subtable}
\end{table}

\begin{example} \label{Examp3}
The assignment matrices for the graph shown in Figure~\ref{Fig4} are given in Table~\ref{Table2}. Therefore,
\[
\mathit{AFD} = 100 \cdot \left( 2 \cdot \left\lvert \frac{2}{3} - \frac{23}{36} \right\rvert + 4 \cdot \left\lvert \frac{1}{3} - \frac{13}{36} \right\rvert + \left\lvert \frac{1}{3} - \frac{5}{18} \right\rvert \right) \cdot \frac{1}{7} = \frac{100}{7} \cdot \frac{8}{36} = 3 \frac{11}{63}.
\]
The maximal distortion occurs for nodes $c$ and $C$:
\[
\mathit{MFD} = 100 \cdot \left\lvert \frac{1}{3} - \frac{5}{18} \right\rvert = 5\frac{5}{9}.
\]
\end{example}

\begin{remark} \label{Rem1}
Section~\ref{Sec56} will present a graph corresponding to a possible UEFA Champions League season ($n=8$), for which maximal fairness distortion exceeds the value derived in Example~\ref{Examp3}.
\end{remark}

Although Proposition~\ref{Prop1} is mainly a theoretical result, it might have practical relevance. For example, the play-offs of the 2024 UEFA European Championship qualifying tournament have determined the last three national teams that can play in the European Championship. While the play-offs contain three independent paths with four participants each, an alternative structure could be to organise two knockout rounds with pairing constraints. Then the assignment in the second knockout round may be given in Figure~\ref{Fig4}, when all the four draw mechanisms have the same level of unfairness.

\subsection{The case of eight nodes: Draw of quarterfinals} \label{Sec52}

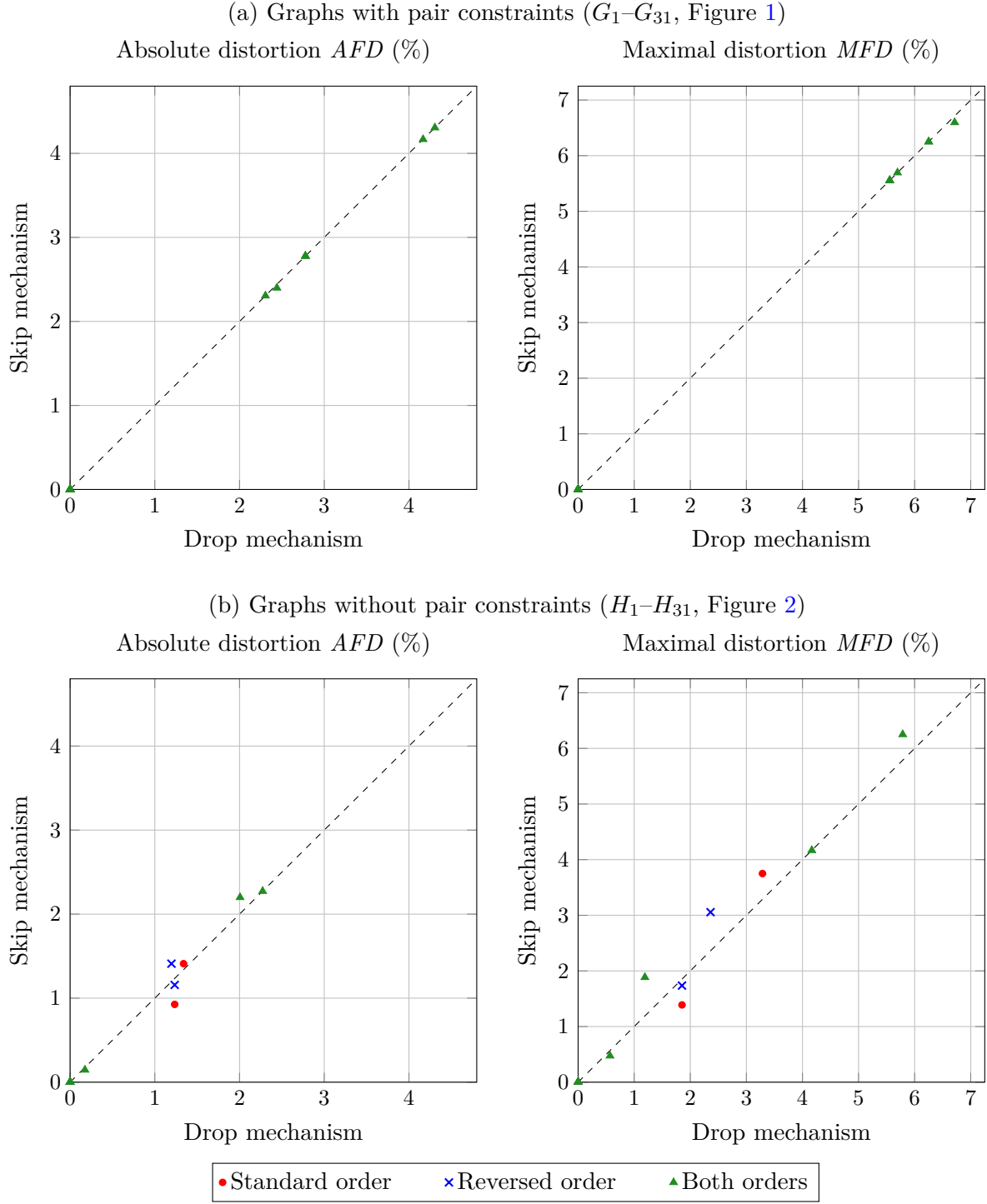
\begin{figure}[t!]
\centering

\begin{subfigure}{\textwidth}
\caption{Graphs with pair constraints ($G_1$--$G_{31}$, Figure~\ref{Fig1})}
\label{Fig5a}

\begin{tikzpicture}
\begin{axis}[
name = axis1,
title = {Absolute distortion $\mathit{AFD}$ (\%)},
title style = {font=\small},
xlabel = $\mathit{AFD}$ of the Drop mechanism,
x label style = {font=\small},
ylabel = $\mathit{AFD}$ of the Skip mechanism,
y label style = {font=\small},
width = 0.5\textwidth,
height = 0.5\textwidth,
nodes near coords,
xmajorgrids = true,
ymajorgrids = true,
xmin = 0,
xmax = 4.8,
ymin = 0,
ymax = 4.8,
]
\addplot [scatter,red,only marks,mark size=1.5pt,point meta=explicit symbolic] coordinates {
};
\addplot [scatter,blue,only marks,mark=x,mark size=2.5pt,thick,point meta=explicit symbolic] coordinates {
};
\addplot [scatter,ForestGreen,only marks,mark=triangle*,mark size=2pt,point meta=explicit symbolic] coordinates {
(0,0)
(2.44107744107744,2.3989898989899)
(0,0)
(4.30555555555557,4.30555555555556)
(2.30555555555551,2.30555555555556)
(0,0)
(0,0)
(4.16666666666667,4.16666666666667)
(0,0)
(0,0)
(2.7777777777778,2.77777777777778)
(0,0)
(2.7777777777778,2.77777777777778)
(0,0)
(0,0)
(0,0)
(0,0)
(0,0)
(0,0)
(2.7777777777778,2.77777777777778)
(0,0)
(0,0)
(0,0)
(0,0)
(0,0)
(0,0)
(0,0)
};
\draw [black,dashed] (rel axis cs:0,0) -- (rel axis cs:1,1);
\end{axis}

\begin{axis}[
at = {(axis1.south east)},
xshift = 0.1\textwidth,
title = {Maximal distortion $\mathit{MFD}$ (\%)},
title style = {font=\small},
xlabel = $\mathit{MFD}$ of the Drop mechanism,
x label style = {font=\small},
ylabel = $\mathit{MFD}$ of the Skip mechanism,
y label style = {font=\small},
width = 0.5\textwidth,
height = 0.5\textwidth,
nodes near coords,
xmajorgrids = true,
ymajorgrids = true,
xmin = 0,
xmax = 7.25,
ymin = 0,
ymax = 7.25,
]
\addplot [scatter,red,only marks,mark size=1.5pt,point meta=explicit symbolic] coordinates {
};
\addplot [scatter,blue,only marks,mark=x,mark size=2.5pt,thick,point meta=explicit symbolic] coordinates {
};
\addplot [scatter,ForestGreen,only marks,mark=triangle*,mark size=2pt,point meta=explicit symbolic] coordinates {
(0,0)
(6.71296296296298,6.59722222222223)
(0,0)
(6.25000000000006,6.25)
(5.69444444444444,5.69444444444444)
(0,0)
(0,0)
(6.25,6.25)
(0,0)
(0,0)
(5.55555555555556,5.55555555555555)
(0,0)
(5.55555555555556,5.55555555555555)
(0,0)
(0,0)
(0,0)
(0,0)
(0,0)
(0,0)
(5.55555555555556,5.55555555555555)
(0,0)
(0,0)
(0,0)
(0,0)
(0,0)
(0,0)
(0,0)
};
\draw [black,dashed] (rel axis cs:0,0) -- (rel axis cs:1,1);
\end{axis}
\end{tikzpicture}
\end{subfigure}

\vspace{0.25cm}
\begin{subfigure}{\textwidth}
\caption{Graphs without pair constraints ($H_1$--$H_{31}$, Figure~\ref{Fig2})}
\label{Fig5b}

\begin{tikzpicture}
\begin{axis}[
name = axis1,
title = {Absolute distortion $\mathit{AFD}$ (\%)},
title style = {font=\small},
xlabel = $\mathit{AFD}$ of the Drop mechanism,
x label style = {font=\small},
ylabel = $\mathit{AFD}$ of the Skip mechanism,
y label style = {font=\small},
width = 0.5\textwidth,
height = 0.5\textwidth,
nodes near coords,
xmajorgrids = true,
ymajorgrids = true,
xmin = 0,
xmax = 4.8,
ymin = 0,
ymax = 4.8,
]
\addplot [scatter,red,only marks,mark size=1.5pt,point meta=explicit symbolic] coordinates {
(1.33903133903132,1.41025641025641)
(1.23456790123459,0.925925925925925)
};
\addplot [scatter,blue,only marks,mark=x,mark size=2.5pt,thick,point meta=explicit symbolic] coordinates {
(1.19658119658119,1.41025641025641)
(1.23456790123457,1.15740740740741)
};
\addplot [scatter,ForestGreen,only marks,mark=triangle*,mark size=2pt,point meta=explicit symbolic] coordinates {
(0,0)
(0,0)
(0.680272108843538,10.7709750566894)
(0.174825174825178,0.145687645687649)
(0,0)
(0,0)
(0,0)
(2.00617283950617,2.19907407407407)
(2.2727272727273,2.27272727272727)
(0,0)
(0,0)
(0,0)
(0,0)
(0,0)
(0,0)
(0,0)
(0,0)
(0,0)
};
\draw [black,dashed] (rel axis cs:0,0) -- (rel axis cs:1,1);
\end{axis}

\begin{axis}[
at = {(axis1.south east)},
xshift = 0.1\textwidth,
title = {Maximal distortion $\mathit{MFD}$ (\%)},
title style = {font=\small},
xlabel = $\mathit{MFD}$ of the Drop mechanism,
x label style = {font=\small},
ylabel = $\mathit{MFD}$ of the Skip mechanism,
y label style = {font=\small},
width = 0.5\textwidth,
height = 0.5\textwidth,
nodes near coords,
xmajorgrids = true,
ymajorgrids = true,
xmin = 0,
xmax = 7.25,
ymin = 0,
ymax = 7.25,
legend style = {font=\small,at={(-0.9,-0.2)},anchor=north west,legend columns=3},
legend entries = {Standard order$\qquad$,Reversed order$\qquad$,Both orders}
]
\addplot [scatter,red,only marks,mark size=1.5pt,point meta=explicit symbolic] coordinates {
(3.28703703703703,3.75)
(1.85185185185196,1.38888888888889)
};
\addplot [scatter,blue,only marks,mark=x,mark size=2.5pt,thick,point meta=explicit symbolic] coordinates {
(2.36111111111115,3.05555555555556)
(1.85185185185187,1.73611111111111)
};
\addplot [scatter,ForestGreen,only marks,mark=triangle*,mark size=2pt,point meta=explicit symbolic] coordinates {
(0,0)
(0,0)
(1.19047619047619,1.88492063492063)
(0.568181818181826,0.473484848484848)
(0,0)
(0,0)
(0,0)
(5.78703703703703,6.25)
(4.1666666666668,4.16666666666667)
(0,0)
(0,0)
(0,0)
(0,0)
(0,0)
(0,0)
(0,0)
(0,0)
(0,0)
};
\draw [black,dashed] (rel axis cs:0,0) -- (rel axis cs:1,1);
\end{axis}
\end{tikzpicture}
\end{subfigure}

\caption{Fairness distortions of the draw mechanisms \\ for balanced bipartite graphs with eight nodes ($n=4$)}
\label{Fig5}
\end{figure}


Figure~\ref{Fig5} presents unfairness for all balanced bipartite graphs with eight nodes such that no node could have the same type as its pair: Figure~\ref{Fig5a} if the pair constraints are effective (graphs in Figure~\ref{Fig1}), while Figure~\ref{Fig5b} if there are only type constraints (graphs in Figure~\ref{Fig2}).

In the presence of pair constraints (Figure~\ref{Fig5a}), both the Drop and Skip mechanisms are fair for 20 graphs and unfair for seven graphs ($G_2$, $G_4$, $G_5$, $G_8$, $G_{11}$, $G_{13}$, $G_{20}$). The absolute and maximal differences are quite high, the probability of matching two nodes can differ by at least five percentage points for all these seven graphs. The ranking of the graphs by the two measures is different, the worst case is $G_4$ for $\mathit{AFD}$ but $G_2$ for $\mathit{MFD}$.

\begin{theorem} \label{Theo1}
For a balanced bipartite graph with eight nodes, pair and type constraints such that pair and type constraints cannot coincide:
\begin{itemize}
\item
The two versions (Standard and Reversed) of the Drop and Skip mechanisms, respectively, coincide;
\item
The Skip mechanism dominates the Drop mechanism as it is less distorted for graph $G_2$.
\end{itemize}
\end{theorem}

The bipartite graphs considered in Theorem~\ref{Theo1} describe all possible scenarios in a hypothetical draw of quarterfinals according to the rules of the UEFA Champions League. Therefore, it is highly relevant for tournament organisers: any such draw should be carried out by the Skip mechanism.

If pair constraints are removed (Figure~\ref{Fig5b}), then both the Drop and Skip mechanisms are fair for 14 graphs and unfair for the remaining six graphs ($H_{3-5}$, $H_{6-8}$, $H_{12-14}$, $H_{15-16}$, $H_{18-20}$, $H_{21-22}$). Since the Drop mechanism is less unfair for graph $H_{18-20}$ and the Skip mechanism is less unfair for graph $H_{6-8}$, neither of them is preferred to the other. However, the Reversed Drop mechanism is fairer than the Standard Drop mechanism. The absolute and maximal distortions are generally lower compared to the graphs with pair constraints.

\begin{result} \label{Result1}
For a balanced bipartite graph with eight nodes and type constraints such that no node has the same type as its pair, the Reversed Drop mechanism is fairer than the Standard Drop mechanism according to both measures $\mathit{AFD}$ and $\mathit{MFD}$ due to graph $H_{12-14}$.
\end{result}


There is no analogous relation for the Skip mechanism: the Reversed Skip mechanism is less unfair for graph $H_{12-14}$ but it is worse for graph $H_{15-16}$.

\subsection{Increasing the number of nodes} \label{Sec53}

\begin{figure}[t!]
\centering

\begin{subfigure}{\textwidth}
\caption{Graphs with pair constraints ($G_1$--$G_{31}$ plus one pair of nodes, Figure~\ref{Fig1})}
\label{Fig6a}

\begin{tikzpicture}
\begin{axis}[
name = axis1,
title = {Absolute distortion $\mathit{AFD}$ (\%)},
title style = {font=\small},
xlabel = $\mathit{AFD}$ of the Drop mechanism,
x label style = {font=\small},
ylabel = $\mathit{AFD}$ of the Skip mechanism,
y label style = {font=\small},
width = 0.5\textwidth,
height = 0.5\textwidth,
nodes near coords,
xmajorgrids = true,
ymajorgrids = true,
xmin = 0,
xmax = 4.1,
ymin = 0,
ymax = 4.1,
]
\addplot [scatter,red,only marks,mark size=1.5pt,point meta=explicit symbolic] coordinates {
(0.575196408529747,0.445566778900112)
(0.838779956427026,0.552287581699346)
(1.32988380537401,1.40359477124183)
(1.08035371011149,0.939830834294502)
(1.74300044091711,1.53968253968254)
(1.89670138888888,1.68923611111111)
(1.63966049382716,1.65625)
(2.63558201058201,2.65277777777778)
(1.13960113960113,0.854700854700854)
(1.5354938271605,1.45833333333333)
};
\addplot [scatter,blue,only marks,mark=x,mark size=2.5pt,thick,point meta=explicit symbolic] coordinates {
(0.588056490834269,0.655443322109989)
(0.774328249818459,0.679738562091503)
(1.23275236020334,1.40114379084967)
(1.10985625186893,1.08852364475202)
(1.70345568783069,1.67162698412698)
(1.69959766313932,1.68576388888889)
(1.51331018518517,1.60069444444444)
(2.89021164021163,2.79960317460317)
(1.13960113960115,1.06837606837607)
(1.8827160493827,1.86342592592592)
};
\addplot [scatter,ForestGreen,only marks,mark=triangle*,mark size=2pt,point meta=explicit symbolic] coordinates {
(0,0)
(0.234250398724091,0.416267942583732)
(0,0)
(0.400891632373112,0.521604938271604)
(0.791340218423561,0.999287749287749)
(0.787037037037033,0.702614379084968)
(1.11082444673777,0.985552115583075)
(0.769562454611469,0.793300653594771)
(0,0)
(0,0)
(0.160108024691384,0.13888888888889)
(2.1728570426487,1.93292297979798)
(2.82366071428573,2.93799603174603)
(3.03780864197532,2.925)
(3.72727272727272,3.59393939393939)
(0.798611111111113,0.861111111111111)
(1.73960170487948,1.77380952380953)
(0.6878306878307,0.785714285714285)
(0,0)
(0.162337662337664,0.135281385281388)
(0,0)
};
\draw [black,dashed] (rel axis cs:0,0) -- (rel axis cs:1,1);
\end{axis}

\begin{axis}[
at = {(axis1.south east)},
xshift = 0.1\textwidth,
title = {Maximal distortion $\mathit{MFD}$ (\%)},
title style = {font=\small},
xlabel = $\mathit{MFD}$ of the Drop mechanism,
x label style = {font=\small},
ylabel = $\mathit{MFD}$ of the Skip mechanism,
y label style = {font=\small},
width = 0.5\textwidth,
height = 0.5\textwidth,
nodes near coords,
xmajorgrids = true,
ymajorgrids = true,
xmin = 0,
xmax = 10.25,
ymin = 0,
ymax = 10.25,
legend style = {font=\small,at={(-0.9,-0.2)},anchor=north west,legend columns=3},
legend entries = {Standard order$\qquad$,Reversed order$\qquad$,Both orders}
]
\addplot [scatter,red,only marks,mark size=1.5pt,point meta=explicit symbolic] coordinates {
(1.2941919191919,1.00252525252525)
(2.68518518518526,1.77777777777778)
(3.51273148148151,3.41666666666667)
(2.26307189542483,2.07516339869281)
(4.84347442680776,4.10119047619048)
(4.71781305114639,4.28571428571429)
(2.50000000000009,2.58333333333334)
(5.83333333333331,5.91666666666667)
(1.85185185185186,1.38888888888889)
(3.5185185185185,3.40277777777778)
};
\addplot [scatter,blue,only marks,mark=x,mark size=2.5pt,thick,point meta=explicit symbolic] coordinates {
(1.32312710437707,1.47474747474748)
(2.45370370370376,2.25)
(3.33333333333335,3.42361111111111)
(2.37495461147419,2.32516339869281)
(4.87819664903005,4.65674603174603)
(3.91644620811283,3.95238095238095)
(2.29166666666673,2.44444444444444)
(6.24999999999996,6.11111111111111)
(1.85185185185188,1.73611111111111)
(4.44444444444441,4.44444444444444)
};
\addplot [scatter,ForestGreen,only marks,mark=triangle*,mark size=2pt,point meta=explicit symbolic] coordinates {
(0,0)
(1.11268939393942,1.97727272727273)
(0,0)
(1.17206790123456,1.20833333333333)
(1.22188093542263,1.99252136752137)
(1.6820987654321,1.66666666666667)
(2.36050194931776,2.09429824561403)
(1.73032407407407,1.5625)
(0,0)
(0,0)
(0.320216049382771,0.277777777777777)
(4.48565516273854,4.11489898989899)
(7.52976190476193,7.65476190476191)
(5.96836419753091,5.73611111111111)
(9.31818181818183,8.98484848484848)
(0.879629629629657,1.09722222222223)
(2.88111772486773,2.75496031746032)
(1.2037037037037,1.375)
(0,0)
(0.568181818181798,0.473484848484848)
(0,0)
};
\draw [black,dashed] (rel axis cs:0,0) -- (rel axis cs:1,1);
\end{axis}
\end{tikzpicture}
\end{subfigure}

\vspace{0.25cm}
\begin{subfigure}{\textwidth}
\caption{Graphs without pair constraints ($H_1$--$H_{31}$ plus one pair of nodes, Figure~\ref{Fig2})}
\label{Fig6b}

\begin{tikzpicture}
\begin{axis}[
name = axis1,
title = {Absolute distortion $\mathit{AFD}$ (\%)},
title style = {font=\small},
xlabel = $\mathit{AFD}$ of the Drop mechanism,
x label style = {font=\small},
ylabel = $\mathit{AFD}$ of the Skip mechanism,
y label style = {font=\small},
width = 0.5\textwidth,
height = 0.5\textwidth,
nodes near coords,
xmajorgrids = true,
ymajorgrids = true,
xmin = 0,
xmax = 4.1,
ymin = 0,
ymax = 4.1,
]
\addplot [scatter,red,only marks,mark size=1.5pt,point meta=explicit symbolic] coordinates {
(0.414141414141424,0.666666666666666)
(0.256613756613751,0.412698412698412)
(0.624338624338617,0.761904761904762)
(0.789566508864749,0.892230576441104)
(0.679138321995467,0.734693877551021)
(1.32638888888888,1.56666666666667)
};
\addplot [scatter,blue,only marks,mark=x,mark size=2.5pt,thick,point meta=explicit symbolic] coordinates {
(0.383838383838391,0.666666666666666)
(0.240740740740732,0.417989417989418)
(0.550264550264545,0.761904761904762)
(0.705096073517122,0.892230576441103)
(0.544217687074826,0.734693877551021)
(0.955555555555559,1.26666666666667)
};
\addplot [scatter,ForestGreen,only marks,mark=triangle*,mark size=2pt,point meta=explicit symbolic] coordinates {
(0,0)
(0,0)
(0.180602006688967,0.354515050167223)
(0.106060606060613,0.287878787878788)
(0.0916442048517484,0.155136268343816)
(0,0)
(0.790123456790121,1.14285714285714)
(0.451929012345677,0.625)
(0,0)
(0.622222222222225,0.755555555555556)
(0.80495219530306,0.898078529657478)
(1.29557007988381,1.41176470588235)
(0,0)
(1.99122807017544,2.23684210526316)
};
\draw [black,dashed] (rel axis cs:0,0) -- (rel axis cs:1,1);
\end{axis}

\begin{axis}[
at = {(axis1.south east)},
xshift = 0.1\textwidth,
title = {Maximal distortion $\mathit{MFD}$ (\%)},
title style = {font=\small},
xlabel = $\mathit{MFD}$ of the Drop mechanism,
x label style = {font=\small},
ylabel = $\mathit{MFD}$ of the Skip mechanism,
y label style = {font=\small},
width = 0.5\textwidth,
height = 0.5\textwidth,
nodes near coords,
xmajorgrids = true,
ymajorgrids = true,
xmin = 0,
xmax = 10.25,
ymin = 0,
ymax = 10.25,
legend style = {font=\small,at={(-0.9,-0.2)},anchor=north west,legend columns=3},
legend entries = {Standard order$\qquad$,Reversed order$\qquad$,Both orders}
]
\addplot [scatter,red,only marks,mark size=1.5pt,point meta=explicit symbolic] coordinates {
(1.5555555555556,2.33333333333334)
(0.907407407407429,1.5)
(1.63888888888892,2)
(2.13712522045853,2.38095238095238)
(3.12202380952388,3.64285714285715)
(6.22222222222223,7.16666666666667)
};
\addplot [scatter,blue,only marks,mark=x,mark size=2.5pt,thick,point meta=explicit symbolic] coordinates {
(1.16666666666665,2.16666666666667)
(0.61111111111109,1.38888888888888)
(1.44444444444448,2)
(3.04761904761908,3.26984126984127)
(1.89285714285721,2.89285714285715)
(4.7638888888889,6.16666666666667)
};
\addplot [scatter,ForestGreen,only marks,mark=triangle*,mark size=2pt,point meta=explicit symbolic] coordinates {
(0,0)
(0,0)
(0.51923076923075,1.01923076923077)
(0.145833333333362,0.395833333333334)
(0.48113207547171,0.814465408805032)
(0,0)
(3.32407407407409,4.77777777777778)
(1.87731481481482,2.69444444444444)
(0,0)
(1.05555555555555,1.16666666666667)
(1.90608465608467,2.23015873015873)
(5.50617283950621,6)
(0,0)
(7.85416666666667,8.75)
};
\draw [black,dashed] (rel axis cs:0,0) -- (rel axis cs:1,1);
\end{axis}
\end{tikzpicture}
\end{subfigure}

\caption{Fairness distortions of the draw mechanisms \\ for selected balanced bipartite graphs with 10 nodes ($n=5$)}
\label{Fig6}
\end{figure}


\begin{figure}[t!]
\centering

\begin{subfigure}{\textwidth}
\caption{Graphs with pair constraints ($G_1$--$G_{31}$ plus two pairs of nodes, Figure~\ref{Fig1})}
\label{Fig7a}

\begin{tikzpicture}
\begin{axis}[
name = axis1,
title = {Absolute distortion $\mathit{AFD}$ (\%)},
title style = {font=\small},
xlabel = $\mathit{AFD}$ of the Drop mechanism,
x label style = {font=\small},
ylabel = $\mathit{AFD}$ of the Skip mechanism,
y label style = {font=\small},
width = 0.5\textwidth,
height = 0.5\textwidth,
nodes near coords,
xmajorgrids = true,
ymajorgrids = true,
xmin = 0,
xmax = 1.5,
ymin = 0,
ymax = 1.5,
]
\addplot [scatter,red,only marks,mark size=1.5pt,point meta=explicit symbolic] coordinates {
(0.298701946274677,0.361866077667965)
(0.342065125089788,0.337595532039976)
(0.397361201644333,0.527814209223831)
(0.261620751409832,0.371456332876086)
(0.437208912317014,0.538539930835528)
(0.341191104929326,0.447137782436524)
(0.29047247578992,0.478068007829912)
(0.832881407259646,1.02368560253125)
(0.312910012854697,0.211921215415276)
(0.374111099805549,0.197564780898114)
(0.731993395725481,0.589673436877738)
};
\addplot [scatter,blue,only marks,mark=x,mark size=2.5pt,thick,point meta=explicit symbolic] coordinates {
(0.298704702006573,0.390305231107118)
(0.330551064733154,0.35371350186165)
(0.366202314800251,0.543789289242121)
(0.289396623990251,0.390775034293552)
(0.426135914831561,0.571837651633249)
(0.333679827626382,0.466724677023419)
(0.321499473483577,0.500860030621936)
(0.875307132196234,1.01431715389394)
(0.322723684551085,0.351153039832286)
(0.342214138741933,0.332417582417582)
(0.649035576795451,0.633191955396256)
};
\addplot [scatter,ForestGreen,only marks,mark=triangle*,mark size=2pt,point meta=explicit symbolic] coordinates {
(0,0)
(0.107650786765948,0.159630352538616)
(0,0)
(0.154251608050623,0.224396008403362)
(0.302345807506964,0.458098857183636)
(0.367949817101051,0.465477823502515)
(0.100520348342374,0.0293697996455505)
(0.243111292858728,0.372543734559678)
(0.0618727195481796,0.115659519168292)
(0.432026175895945,0.46743295019157)
(0.801469782955018,0.746059645253194)
(0.364873088689801,0.384741902834007)
(0.670770227049451,0.949757512055246)
(0.690035579561043,0.848641975308641)
(0.687544165196933,0.909497354497355)
(0.477750079138961,0.341405508072174)
(0.482694911502165,0.617019071310116)
(0.445166338772197,0.28091157678146)
(1.34462129714655,1.33287419651056)
(0.703076774691382,0.538580246913581)
};
\draw [black,dashed] (rel axis cs:0,0) -- (rel axis cs:1,1);
\end{axis}

\begin{axis}[
at = {(axis1.south east)},
xshift = 0.1\textwidth,
title = {Maximal distortion $\mathit{MFD}$ (\%)},
title style = {font=\small},
xlabel = $\mathit{MFD}$ of the Drop mechanism,
x label style = {font=\small},
ylabel = $\mathit{MFD}$ of the Skip mechanism,
y label style = {font=\small},
width = 0.5\textwidth,
height = 0.5\textwidth,
nodes near coords,
xmajorgrids = true,
ymajorgrids = true,
xmin = 0,
xmax = 8.25,
ymin = 0,
ymax = 8.25,
legend style = {font=\small,at={(-0.9,-0.2)},anchor=north west,legend columns=3},
legend entries = {Standard order$\qquad$,Reversed order$\qquad$,Both orders}
]
\addplot [scatter,red,only marks,mark size=1.5pt,point meta=explicit symbolic] coordinates {
(0.580902231830868,0.714033018867924)
(1.56851851851843,1.37962962962963)
(1.36085091532357,1.88008316841014)
(0.467923739711942,0.755401234567901)
(1.39564455703091,1.45384928954111)
(1.25767856102952,1.70619322152341)
(0.550228594524232,1.02535273368606)
(4.25643966599363,5.07544258094572)
(1.05607129338458,0.715234102026555)
(1.11729772927691,0.642085537918871)
(2.21436720098232,1.72253335324571)
};
\addplot [scatter,blue,only marks,mark=x,mark size=2.5pt,thick,point meta=explicit symbolic] coordinates {
(0.545785290006939,0.73833857442348)
(1.41296296296292,1.46527777777778)
(1.08409551408905,1.77900292149656)
(0.536731610082294,0.799768518518518)
(1.27695242740119,1.6498369438621)
(0.95313252193493,1.60974260423946)
(0.54203672209629,0.995260141093474)
(4.28125533814724,4.95661542045189)
(1.08919243535988,1.18514150943396)
(1.09261923133452,1.08035714285714)
(1.98221378020048,1.88457039028275)
};
\addplot [scatter,ForestGreen,only marks,mark=triangle*,mark size=2pt,point meta=explicit symbolic] coordinates {
(0,0)
(0.780468204053092,1.15732005590496)
(0,0)
(0.651308324255606,0.814406318082789)
(0.773765599354204,1.23889870120602)
(0.754019694297536,1.08553791887125)
(0.226170783770452,0.0660820492024872)
(0.651640053697788,0.964625521804779)
(0.20108633853152,0.375893437296945)
(1.40408507166182,1.51915708812261)
(2.6047767946038,2.42469384707288)
(1.47520796106778,1.68433235867446)
(3.69871037096964,4.77421191417955)
(1.69331490054872,2.0462962962963)
(3.66999559082877,4.70271164021164)
(1.19652777777783,0.840277777777779)
(1.31653972933688,1.24675234936429)
(1.33549901631655,0.829536224821312)
(3.697708567153,3.66540404040404)
(2.10923032407412,1.61574074074074)
};
\draw [black,dashed] (rel axis cs:0,0) -- (rel axis cs:1,1);
\end{axis}
\end{tikzpicture}
\end{subfigure}

\vspace{0.25cm}
\begin{subfigure}{\textwidth}
\caption{Graphs without pair constraints ($H_1$--$H_{31}$ plus two pairs of nodes, Figure~\ref{Fig2})}
\label{Fig7b}

\begin{tikzpicture}
\begin{axis}[
name = axis1,
title = {Absolute distortion $\mathit{AFD}$ (\%)},
title style = {font=\small},
xlabel = $\mathit{AFD}$ of the Drop mechanism,
x label style = {font=\small},
ylabel = $\mathit{AFD}$ of the Skip mechanism,
y label style = {font=\small},
width = 0.5\textwidth,
height = 0.5\textwidth,
nodes near coords,
xmajorgrids = true,
ymajorgrids = true,
xmin = 0,
xmax = 1.5,
ymin = 0,
ymax = 1.5,
]
\addplot [scatter,red,only marks,mark size=1.5pt,point meta=explicit symbolic] coordinates {
(0.139644814154627,0.279263220439691)
(0.118137039697745,0.263150542784163)
(0.269356261022915,0.47089947089947)
(0.474759945130307,0.745679012345679)
(0.254825498575496,0.405982905982905)
(0.490697295536005,0.74356467904855)
};
\addplot [scatter,blue,only marks,mark=x,mark size=2.5pt,thick,point meta=explicit symbolic] coordinates {
(0.132237406747227,0.279263220439691)
(0.111713428586646,0.263729246487866)
(0.247134038800716,0.47089947089947)
(0.452674897119323,0.74567901234568)
(0.216168091168065,0.405982905982905)
(0.446790485500152,0.74356467904855)
};
\addplot [scatter,ForestGreen,only marks,mark=triangle*,mark size=2pt,point meta=explicit symbolic] coordinates {
(0,0)
(0,0)
(0.0597572362278245,0.138188608776844)
(0.069042538056659,0.182173851187936)
(0.0618727195481796,0.142201248209536)
(0,0)
(0.29295634920635,0.543650793650794)
(0.181831054471436,0.335324571883711)
(0,0)
(0.268646526711035,0.481310803891449)
(0.284832481190496,0.501533857089412)
(0.735336283094566,1.10257124292212)
(0,0)
(0.771574074074071,1.06296296296296)
};
\draw [black,dashed] (rel axis cs:0,0) -- (rel axis cs:1,1);
\end{axis}

\begin{axis}[
at = {(axis1.south east)},
xshift = 0.1\textwidth,
title = {Maximal distortion $\mathit{MFD}$ (\%)},
title style = {font=\small},
xlabel = $\mathit{MFD}$ of the Drop mechanism,
x label style = {font=\small},
ylabel = $\mathit{MFD}$ of the Skip mechanism,
y label style = {font=\small},
width = 0.5\textwidth,
height = 0.5\textwidth,
nodes near coords,
xmajorgrids = true,
ymajorgrids = true,
xmin = 0,
xmax = 8.25,
ymin = 0,
ymax = 8.25,
legend style = {font=\small,at={(-0.9,-0.2)},anchor=north west,legend columns=3},
legend entries = {Standard order$\qquad$,Reversed order$\qquad$,Both orders}
]
\addplot [scatter,red,only marks,mark size=1.5pt,point meta=explicit symbolic] coordinates {
(0.752505446623097,1.3562091503268)
(0.355034057045608,0.669540229885057)
(0.53871252204587,0.941798941798944)
(0.939943415637956,1.62962962962963)
(1.63468660968671,2.35042735042735)
(3.51982323232322,5.01010101010101)
};
\addplot [scatter,blue,only marks,mark=x,mark size=2.5pt,thick,point meta=explicit symbolic] coordinates {
(0.578431372549021,1.30065359477124)
(0.257503192848008,0.637132822477648)
(0.49426807760152,0.941798941798944)
(1.26311728395064,1.80555555555556)
(1.01709401709393,2.12820512820513)
(2.77954545454538,4.70454545454546)
};
\addplot [scatter,ForestGreen,only marks,mark=triangle*,mark size=2pt,point meta=explicit symbolic] coordinates {
(0,0)
(0,0)
(0.253968253968254,0.587301587301586)
(0.0949334898278864,0.250489045383412)
(0.20108633853152,0.28440249641907)
(0,0)
(1.6712301587302,2.96428571428571)
(1.07268518518517,1.93209876543209)
(0,0)
(0.680224867724932,1.00198412698412)
(0.588566997131759,0.889450056116722)
(1.28683849541548,1.92949967511371)
(0,0)
(5.52638888888876,7.66666666666667)
};
\draw [black,dashed] (rel axis cs:0,0) -- (rel axis cs:1,1);
\end{axis}
\end{tikzpicture}
\end{subfigure}

\caption{Fairness distortions of the draw mechanisms \\ for selected balanced bipartite graphs with 12 nodes ($n=6$)}
\label{Fig7}
\end{figure}


In order to get more insight into the draw mechanisms, Figures~\ref{Fig6} and \ref{Fig7} uncover their unfairness if one pair and two pairs of nodes are added (without further type constraints), respectively, to the previous graphs with $n=4$.

Under pair constraints, the Drop and Skip mechanisms are fair for six graphs if $n=5$ but only for two graphs if $n=6$. Neither the Drop nor the Skip mechanism is better than the other. The draw order has a surprising effect: for graph $G_{11}$ and $\mathit{MFD}$, the Reversed Drop mechanism has a higher unfairness by 2.2\% compared to the Standard version if $n=5$, but a lower unfairness by 6\% if $n=6$. Thus, merely adding a pair of nodes without type constraints can change the optimal draw order. Analogously, for graph $G_{26}$, the Reversed Skip mechanism is less fair according to both measures than the Standard Skip mechanism if $n=5$, but their order is changed if $n=6$.
Maximal unfairness exceeds 8(\%) for graph $G_{22}$ if $n=5$, which indicates a quite serious problem. However, the absolute distortions are not worse than in Example~\ref{Examp1}.

Adding a pair of nodes without type constraints does not necessarily improve fairness. For example, the unfairness of our mechanisms for graph $G_{18}$ is at least quintupled if $n=6$ compared to $n=5$, although the number of valid assignments increases from 10 to 93.

For the graphs without pair constraints (Figures~\ref{Fig6b} and \ref{Fig7b}), the Drop and Skip mechanisms are fair for five and unfair for the remaining 15 graphs. Among the latter, the Reversed Drop mechanism is usually better than the Standard version, for instance, its average fairness distortion is the same for nine graphs and is lower for six graphs. However, in the case of graph $H_{26}$, $\mathit{MFD}$ favours the Standard Drop mechanism for both $n=5$ and $n=6$. Similarly, the Reversed Skip mechanism is usually less unfair than the Standard Skip mechanism.

Regarding the two main draw mechanisms, we have a more robust finding.

\begin{proposition} \label{Prop2}
For a balanced bipartite graph with 10 or 12 nodes, and type constraints affecting at most four pairs of nodes such that no node has the same type as its pair, the Reversed (Standard) Drop mechanism is fairer than the Reversed (Standard) Skip mechanism according to both measures $\mathit{AFD}$ and $\mathit{MFD}$.
\end{proposition}

In the examples without pair constraints, having an additional unconstrained pair of nodes generally reduces the level of unfairness, except if the draw mechanisms are fair only due to the symmetry of the underlying graph such as $H_{9-10}$. However, this cannot explain the higher maximal fairness distortion of the Standard Skip mechanism for graph $H_{15-16}$ if $n$ is increased from 4 to 5.

Figures~\ref{Fig5}--\ref{Fig7} suggest an interesting pattern on the impact of the number of nodes $n$.

\begin{result} \label{Result2}
If type constraints affect at most four pairs of nodes such that no node has the same type as its pair, then a higher number of nodes increases the advantage of the Drop mechanism with respect to fairness, independently of the presence of pair constraints.
\end{result}

According to Result~\ref{Result2}, having more nodes without additional constraints tends to reduce the unfairness of the Drop mechanism more rapidly than the unfairness of the Skip mechanism.

\subsection{Sufficient conditions for fairness} \label{Sec54}

The calculations in Sections~\ref{Sec52} and \ref{Sec53} allow to formulate sufficient conditions for the fairness of the draw procedures.

\begin{proposition} \label{Prop3}
The Drop and the Skip mechanisms are fair if one of the following conditions holds:
\begin{itemize}
\item
There is only one valid matching (graphs $G_{18}$, $G_{19}$, $G_{21}$, $G_{23}$--$G_{27}$ if $n=4$);
\item
There are no type constraints (graphs $G_1$ and $H_1$);
\item
There are no pair constraints and the type constraints affect only one node on one side of the bipartite graph (graphs $H_2$, $H_{11}$, $H_{28}$);
\item
The balanced bipartite graph can be decomposed into two unconnected balanced bipartite graphs such that the complement of the first subgraph is complete and the draw mechanism is fair for the second subgraph (graphs $G_3$ and $H_{23}$).
\end{itemize}
\end{proposition}

\subsection{Graphs given by the UEFA Champions League Round of 16} \label{Sec55}

\input{Figure8_UEFA_CL_Drop_Skip}

Figure~\ref{Fig8} compares the distortions of the Drop and Skip mechanisms for the 21 graphs given by the historical UEFA Champions League seasons between 2003/04 and 2023/24. The officially used Drop mechanism is fairer in all cases according to $\mathit{AFD}$. This robust result almost holds with respect to maximal fairness distortion, too, except for the Standard Drop and Standard Skip mechanisms in the 2014/15 season (where the Reversed Drop is the east biased).

\begin{table}[t!]
  \centering
  \caption{Fairness distortions for the 2022/23 UEFA Champions League Round of 16 draw}
  \label{Table3}
  
\begin{subtable}{\textwidth}
\centering
  \caption{Standard Drop mechanism}
  \label{Table3a}
\begin{footnotesize}
    \begin{tabularx}{0.8\textwidth}{l CCCC CCCC} \toprule
          & V1    & V2    & V3    & V4    & V5    & V6    & V7    & V8 \\    \bottomrule
    U1    & X     & \cellcolor{ForestGreen!1.67938445820437} 0.08 & X     & \cellcolor{red!0.948637564499477} 0.05 & X     & \cellcolor{red!1.57991378740968} 0.08 & \cellcolor{red!0.855217564499522} 0.04 & \cellcolor{ForestGreen!1.70438445820437} 0.09 \\
    U2    & \cellcolor{ForestGreen!20.7092020227038} 1.04 & X     & \cellcolor{red!4.53824833849326} 0.23 & \cellcolor{red!1.61910740970073} 0.08 & \cellcolor{red!6.56388392156865} 0.33 & \cellcolor{ForestGreen!1.93813351909189} 0.1 & \cellcolor{red!1.75032740970071} 0.09 & \cellcolor{red!8.17576846233231} 0.41 \\
    U3    & \cellcolor{red!54.6364422910217} 2.73 & \cellcolor{ForestGreen!27.0181978328173} 1.35 & X     & X     & \cellcolor{ForestGreen!0.618506625386983} 0.03 & X     & X     & \cellcolor{ForestGreen!26.9997378328173} 1.35 \\
    U4    & X     & \cellcolor{red!6.30230961816308} 0.32 & \cellcolor{ForestGreen!0.0465216718266714} 0 & X     & \cellcolor{ForestGreen!9.93674986584109} 0.5 & \cellcolor{ForestGreen!0.394022249741999} 0.02 & \cellcolor{ForestGreen!2.18878544891643} 0.11 & \cellcolor{red!6.26376961816305} 0.31 \\
    U5    & X     & \cellcolor{red!0.700430897832838} 0.04 & \cellcolor{ForestGreen!8.00744864809083} 0.4 & \cellcolor{red!1.7780545510836} 0.09 & X     & \cellcolor{red!3.07311775025798} 0.15 & \cellcolor{red!1.79383455108362} 0.09 & \cellcolor{red!0.662010897832799} 0.03 \\
    U6    & \cellcolor{ForestGreen!13.300418245614} 0.67 & \cellcolor{red!7.37904369453044} 0.37 & \cellcolor{ForestGreen!0.902484685242511} 0.05 & \cellcolor{ForestGreen!3.91124148606808} 0.2 & \cellcolor{red!7.29575851393188} 0.36 & X     & \cellcolor{ForestGreen!3.92928148606808} 0.2 & \cellcolor{red!7.36862369453042} 0.37 \\
    U7    & X     & \cellcolor{red!6.2229696181631} 0.31 & \cellcolor{ForestGreen!0.0406216718266683} 0 & \cellcolor{ForestGreen!2.20260544891643} 0.11 & \cellcolor{ForestGreen!9.8629098658411} 0.49 & \cellcolor{ForestGreen!0.350782249741999} 0.02 & X     & \cellcolor{red!6.23394961816309} 0.31 \\
    U8    & \cellcolor{ForestGreen!20.6268220227038} 1.03 & \cellcolor{red!8.09282846233231} 0.4 & \cellcolor{red!4.45882833849326} 0.22 & \cellcolor{red!1.76804740970071} 0.09 & \cellcolor{red!6.55852392156864} 0.33 & \cellcolor{ForestGreen!1.97009351909189} 0.1 & \cellcolor{red!1.71868740970071} 0.09 & X \\ \toprule
    \end{tabularx}
\end{footnotesize}
\end{subtable}

\vspace{0.25cm}
\begin{subtable}{\textwidth}
\centering
  \caption{Reversed Drop mechanism}
  \label{Table3b}
\begin{footnotesize}
    \begin{tabularx}{0.8\textwidth}{l CCCC CCCC} \toprule
          & V1    & V2    & V3    & V4    & V5    & V6    & V7    & V8 \\    \bottomrule
    U1    & X     & \cellcolor{ForestGreen!0.816404458204356} 0.04 & X     & \cellcolor{red!0.551317564499498} 0.03 & X     & \cellcolor{red!0.657873787409669} 0.03 & \cellcolor{red!0.426237564499499} 0.02 & \cellcolor{ForestGreen!0.819024458204365} 0.04 \\
    U2    & \cellcolor{ForestGreen!22.1259020227039} 1.11 & X     & \cellcolor{red!4.38284833849328} 0.22 & \cellcolor{red!2.03046740970075} 0.1 & \cellcolor{red!7.76410392156868} 0.39 & \cellcolor{ForestGreen!2.17331351909189} 0.11 & \cellcolor{red!2.05206740970071} 0.1 & \cellcolor{red!8.06972846233231} 0.4 \\
    U3    & \cellcolor{red!59.0168422910217} 2.95 & \cellcolor{ForestGreen!28.1715978328173} 1.41 & X     & X     & \cellcolor{ForestGreen!2.75814662538698} 0.14 & X     & X     & \cellcolor{ForestGreen!28.0870978328173} 1.4 \\
    U4    & X     & \cellcolor{red!5.96980961816307} 0.3 & \cellcolor{red!0.330678328173351} 0.02 & X     & \cellcolor{ForestGreen!10.5366898658411} 0.53 & \cellcolor{red!0.298457750257997} 0.01 & \cellcolor{ForestGreen!2.0169454489164} 0.1 & \cellcolor{red!5.95468961816309} 0.3 \\
    U5    & X     & \cellcolor{red!1.8045908978328} 0.09 & \cellcolor{ForestGreen!8.66324864809082} 0.43 & \cellcolor{red!1.00171455108361} 0.05 & X     & \cellcolor{red!3.15339775025802} 0.16 & \cellcolor{red!0.958174551083602} 0.05 & \cellcolor{red!1.74537089783283} 0.09 \\
    U6    & \cellcolor{ForestGreen!14.843198245614} 0.74 & \cellcolor{red!7.17346369453042} 0.36 & \cellcolor{ForestGreen!0.934404685242529} 0.05 & \cellcolor{ForestGreen!3.42726148606809} 0.17 & \cellcolor{red!8.35861851393188} 0.42 & X     & \cellcolor{ForestGreen!3.45544148606808} 0.17 & \cellcolor{red!7.12822369453045} 0.36 \\
    U7    & X     & \cellcolor{red!6.01106961816306} 0.3 & \cellcolor{red!0.438438328173363} 0.02 & \cellcolor{ForestGreen!2.12072544891639} 0.11 & \cellcolor{ForestGreen!10.6274498658411} 0.53 & \cellcolor{red!0.290557750257991} 0.01 & X     & \cellcolor{red!6.00810961816306} 0.3 \\
    U8    & \cellcolor{ForestGreen!22.0477420227038} 1.1 & \cellcolor{red!8.02906846233231} 0.4 & \cellcolor{red!4.44568833849329} 0.22 & \cellcolor{red!1.96448740970073} 0.1 & \cellcolor{red!7.79956392156866} 0.39 & \cellcolor{ForestGreen!2.22697351909185} 0.11 & \cellcolor{red!2.03590740970072} 0.1 & X \\ \toprule
    \end{tabularx}
\end{footnotesize}
\end{subtable}

\vspace{0.25cm}
\begin{subtable}{\textwidth}
\centering
  \caption{Standard Skip mechanism}
  \label{Table3c}
\begin{footnotesize}
    \begin{tabularx}{0.8\textwidth}{l CCCC CCCC} \toprule
          & V1    & V2    & V3    & V4    & V5    & V6    & V7    & V8 \\    \bottomrule
    U1    & X     & \cellcolor{ForestGreen!4.09620445820436} 0.2 & X     & \cellcolor{red!3.0095575644995} 0.15 & X     & \cellcolor{red!2.3363737874097} 0.12 & \cellcolor{red!3.0597575644995} 0.15 & \cellcolor{ForestGreen!4.30948445820434} 0.22 \\
    U2    & \cellcolor{ForestGreen!32.0990620227038} 1.6 & X     & \cellcolor{red!5.97122833849328} 0.3 & \cellcolor{red!2.56082740970071} 0.13 & \cellcolor{red!10.0955239215686} 0.5 & \cellcolor{ForestGreen!2.49849351909187} 0.12 & \cellcolor{red!2.58774740970075} 0.13 & \cellcolor{red!13.3822284623323} 0.67 \\
    U3    & \cellcolor{red!88.0465622910217} 4.4 & \cellcolor{ForestGreen!42.3225978328173} 2.12 & X     & X     & \cellcolor{ForestGreen!3.38852662538697} 0.17 & X     & X     & \cellcolor{ForestGreen!42.3354378328173} 2.12 \\
    U4    & X     & \cellcolor{red!8.7380496181631} 0.44 & \cellcolor{red!1.31393832817334} 0.07 & X     & \cellcolor{ForestGreen!14.2824698658411} 0.71 & \cellcolor{ForestGreen!0.944302249742013} 0.05 & \cellcolor{ForestGreen!3.51486544891638} 0.18 & \cellcolor{red!8.6896496181631} 0.43 \\
    U5    & X     & \cellcolor{red!2.44747089783282} 0.12 & \cellcolor{ForestGreen!12.1562486480908} 0.61 & \cellcolor{red!1.54029455108357} 0.08 & X     & \cellcolor{red!4.21189775025799} 0.21 & \cellcolor{red!1.4444745510836} 0.07 & \cellcolor{red!2.51211089783282} 0.13 \\
    U6    & \cellcolor{ForestGreen!23.863378245614} 1.19 & \cellcolor{red!13.3970036945304} 0.67 & \cellcolor{ForestGreen!2.36428468524252} 0.12 & \cellcolor{ForestGreen!6.16830148606812} 0.31 & \cellcolor{red!11.7864585139319} 0.59 & X     & \cellcolor{ForestGreen!6.16778148606811} 0.31 & \cellcolor{red!13.3802836945304} 0.67 \\
    U7    & X     & \cellcolor{red!8.60642961816305} 0.43 & \cellcolor{red!1.29961832817332} 0.06 & \cellcolor{ForestGreen!3.48284544891642} 0.17 & \cellcolor{ForestGreen!14.3355498658411} 0.72 & \cellcolor{ForestGreen!0.768302249742003} 0.04 & X     & \cellcolor{red!8.68064961816306} 0.43 \\
    U8    & \cellcolor{ForestGreen!32.0841220227038} 1.6 & \cellcolor{red!13.2298484623323} 0.66 & \cellcolor{red!5.93574833849325} 0.3 & \cellcolor{red!2.54046740970071} 0.13 & \cellcolor{red!10.1245639215686} 0.51 & \cellcolor{ForestGreen!2.33717351909185} 0.12 & \cellcolor{red!2.59066740970071} 0.13 & X \\ \toprule
    \end{tabularx}
\end{footnotesize}
\end{subtable}

\vspace{0.25cm}
\begin{subtable}{\textwidth}
\centering
  \caption{Reversed Skip mechanism}
  \label{Table3d}
\begin{threeparttable}
\begin{footnotesize}
    \begin{tabularx}{0.8\textwidth}{l CCCC CCCC} \toprule
          & V1    & V2    & V3    & V4    & V5    & V6    & V7    & V8 \\    \bottomrule
    U1    & X     & \cellcolor{ForestGreen!4.14240445820435} 0.21 & X     & \cellcolor{red!3.00409756449949} 0.15 & X     & \cellcolor{red!2.28867378740971} 0.11 & \cellcolor{red!3.00299756449951} 0.15 & \cellcolor{ForestGreen!4.15336445820436} 0.21 \\
    U2    & \cellcolor{ForestGreen!32.9323620227038} 1.65 & X     & \cellcolor{red!6.3956883384933} 0.32 & \cellcolor{red!2.74242740970071} 0.14 & \cellcolor{red!10.1625039215686} 0.51 & \cellcolor{ForestGreen!2.34771351909185} 0.12 & \cellcolor{red!2.66320740970072} 0.13 & \cellcolor{red!13.3162484623323} 0.67 \\
    U3    & \cellcolor{red!88.7670422910216} 4.44 & \cellcolor{ForestGreen!41.1920578328173} 2.06 & X     & X     & \cellcolor{ForestGreen!6.30576662538695} 0.32 & X     & X     & \cellcolor{ForestGreen!41.2692178328173} 2.06 \\
    U4    & X     & \cellcolor{red!8.20808961816305} 0.41 & \cellcolor{red!1.28775832817335} 0.06 & X     & \cellcolor{ForestGreen!13.0641098658411} 0.65 & \cellcolor{ForestGreen!0.971482249742028} 0.05 & \cellcolor{ForestGreen!3.59180544891641} 0.18 & \cellcolor{red!8.13154961816309} 0.41 \\
    U5    & X     & \cellcolor{red!2.17269089783284} 0.11 & \cellcolor{ForestGreen!12.3687886480908} 0.62 & \cellcolor{red!1.75595455108357} 0.09 & X     & \cellcolor{red!4.233917750258} 0.21 & \cellcolor{red!1.89109455108361} 0.09 & \cellcolor{red!2.31513089783281} 0.12 \\
    U6    & \cellcolor{ForestGreen!22.929898245614} 1.15 & \cellcolor{red!13.5540036945304} 0.68 & \cellcolor{ForestGreen!3.01320468524252} 0.15 & \cellcolor{ForestGreen!6.65892148606811} 0.33 & \cellcolor{red!12.1783985139319} 0.61 & X     & \cellcolor{ForestGreen!6.58592148606807} 0.33 & \cellcolor{red!13.4555436945304} 0.67 \\
    U7    & X     & \cellcolor{red!8.08938961816308} 0.4 & \cellcolor{red!1.29041832817334} 0.06 & \cellcolor{ForestGreen!3.60428544891639} 0.18 & \cellcolor{ForestGreen!13.1146898658411} 0.66 & \cellcolor{ForestGreen!0.864942249741985} 0.04 & X     & \cellcolor{red!8.20410961816309} 0.41 \\
    U8    & \cellcolor{ForestGreen!32.9047820227039} 1.65 & \cellcolor{red!13.3102884623323} 0.67 & \cellcolor{red!6.4081283384933} 0.32 & \cellcolor{red!2.76072740970074} 0.14 & \cellcolor{red!10.1436639215686} 0.51 & \cellcolor{ForestGreen!2.33845351909184} 0.12 & \cellcolor{red!2.62042740970075} 0.13 & X \\ \toprule
    \end{tabularx}
\end{footnotesize}
    \begin{tablenotes} \footnotesize
\item
X represents a pair of teams that cannot play in the Round of 16.
\item
The numbers show percentages ($100 \cdot \left( p_{ij} - p_{ij}^M \right)$) rounded to two decimal places.
\item
\textcolor{ForestGreen}{Green} (\textcolor{red}{Red}) colour means that the draw mechanism implies a higher (lower) probability than the Uniform mechanism.
\item
Darker colour indicates a higher value.
    \end{tablenotes}
\end{threeparttable}
\end{subtable}
\end{table}


Unsurprisingly, the level of unfairness strongly depends on the set of constraints, $\mathit{AFD}$ varies between 0.051(\%) and 0.363(\%) for the Drop mechanism and can exceed 0.5(\%) for the Skip mechanism. The difference are even higher for $\mathit{MFD}$, the two disadvantageous outliers are the 2017/18 and the 2022/23 seasons. As an illustration, the distortions in the 2022/23 season are presented in Table~\ref{Table3}. The highest absolute bias affects the match Bayern M\"unchen versus Liverpool, which is above 2.7\% for all draw mechanisms that imply a lower probability than the Uniform draw. It can also be seen that the distortions follow the same patterns, the four mechanisms usually change the probabilities in the same direction and roughly with the same magnitude.

Nonetheless, the biases of the Drop and Skip mechanisms are similar, it is impossible to achieve a fundamental improvement by any draw procedure if the constraints have an unfavourable structure. This finding sheds new light on \citet[Result~2]{BoczonWilson2023}: they would probably have concluded that each of the four field-proven mechanisms comes quantitatively close to a constrained-best.

The situation is more obvious after removing the pair constraints (Figure~\ref{Fig8b}), when the superiority of the Drop mechanism becomes undeniable. Both procedures would have been entirely fair in 2003/04 and 2023/24 due to the third condition of Proposition~\ref{Prop3}. However, the Drop mechanism can decrease the average fairness distortion of the Skip mechanism from about 0.445(\%) to below 0.2(\%) for the graph corresponding to the 2013/14 season, which provides an interesting example because the distortions are much lower---0.057(\%) for the Drop and 0.163(\%) for the Skip---with the eight additional pair constraints. On the other hand, the worst maximal fairness distortions are again associated with the 2017/18 and the 2022/23 seasons as in Figure~\ref{Fig8a}.

\begin{result} \label{Result3}
The Drop mechanism is (almost) always better than the Skip mechanism for the graphs corresponding to the historical draws of the UEFA Champions League Round of 16.
\end{result}

In particular, the cumulated $\mathit{MFD}$ ($\mathit{AFD}$) over the 21 seasons is smaller by 40.5\% (39.7\%) under the Standard Drop mechanism than under the Standard Skip mechanism.

\input{Figure9_UEFA_CL_draw_order}

Figure~\ref{Fig9} focuses on the effect of the draw order. As expected, the difference between the Standard and Reversed versions is more moderated than the difference between the Drop and Skip mechanisms. The Skip mechanism is almost insensitive to the draw order, and removing the pair constraints also diminishes the role of the draw order.

In the presence of draw constraints, the Standard Drop mechanism seems to be better than the Reversed Drop. This is quite remarkable for the 2017/18 season, where the Standard Drop decreases both measures of unfairness by approximately 30\% compared to the Reversed Drop. The issue will be further investigated in Section~\ref{Sec56}.

Note that both the Standard and the Reversed Drop mechanisms have the same chance to be fairer than the other version from a purely mathematical point of view: the assumption in Section~\ref{Sec41} that the degree sequence of the set $U$ is not lexicographically smaller is arbitrary. In other words, if the Reversed Drop is preferred for a balanced bipartite graph but sets $U$ and $V$ are exchanged, then the Standard Drop becomes preferred.

\begin{table}[t!]
  \centering
  \caption{Associations with more than one team \\ in the UEFA Champions League Round of 16}
  \label{Table4}
\centerline{
\begin{threeparttable}
    \rowcolors{1}{gray!20}{}
    \begin{tabularx}{1.2\textwidth}{r CCC CCC CCC CCC CCC CCC} \toprule \hiderowcolors
    \multirow{2}[0]{*}{Season} & \multicolumn{3}{c}{England} & \multicolumn{3}{c}{France} & \multicolumn{3}{c}{Germany} & \multicolumn{3}{c}{Italy} & \multicolumn{3}{c}{Portugal} & \multicolumn{3}{c}{Spain} \\
          & 1    & 2    & $\Delta$  & 1    & 2    & $\Delta$  & 1    & 2    & $\Delta$  & 1    & 2    & $\Delta$  & 1    & 2    & $\Delta$  & 1    & 2    & $\Delta$ \\  \bottomrule \showrowcolors
    2003/04 & 3     & 0     & 3     & 2     & 0     & 2     & 0     & 2     & $-$2    & 2     & 0     & 2     & 0     & 1     & $-$1    & 1     & 3     & $-$2 \\
    2004/05 & 2     & 2     & 0     & 2     & 0     & 2     & 1     & 2     & $-$1    & 3     & 0     & 3     & 0     & 1     & $-$1    & 0     & 2     & $-$2 \\
    2005/06 & 2$^\ast$ & 1     & 0$^\ast$     & 1     & 0     & 1     & 0     & 2     & $-$2    & 3     & 0     & 3     & 0     & 1     & $-$1    & 2     & 1     & 1 \\
    2006/07 & 4     & 0     & 4     & 1     & 1     & 0     & 1     & 0     & 1     & 1     & 2     & $-$1    & 0     & 1     & $-$1    & 1     & 2     & $-$1 \\
    2007/08 & 2     & 2     & 0     & 0     & 1     & $-$1    & 0     & 1     & $-$1    & 2     & 1     & 1     & 1     & 0     & 1     & 3     & 0     & 3 \\
    2008/09 & 2     & 2     & 0     & 0     & 1     & $-$1    & 1     & 0     & 1     & 2     & 1     & 1     & 1     & 1     & 0     & 1     & 3     & $-$2 \\
    2009/10 & 3     & 0     & 3     & 1     & 1     & 0     & 0     & 2     & $-$2    & 1     & 2     & $-$1    & 0     & 1     & $-$1    & 3     & 0     & 3 \\
    2010/11 & 3     & 1     & 2     & 0     & 2     & $-$2    & 2     & 0     & 2     & 0     & 3     & $-$3    & 0     & 0     & 0     & 2     & 1     & 1 \\
    2011/12 & 2     & 0     & 2     & 0     & 2     & $-$2    & 1     & 1     & 0     & 1     & 2     & $-$1    & 1     & 0     & 1     & 2     & 0     & 2 \\
    2012/13 & 1     & 1     & 0     & 1     & 0     & 1     & 3     & 0     & 3     & 1     & 1     & 0     & 0     & 1     & $-$1    & 2     & 2     & 0 \\
    2013/14 & 2     & 2     & 0     & 1     & 0     & 1     & 2     & 2     & 0     & 0     & 1     & $-$1    & 0     & 0     & 0     & 3     & 0     & 3 \\
    2014/15 & 1     & 2     & $-$1    & 1     & 1     & 0     & 2     & 2     & 0     & 0     & 1     & $-$1    & 1     & 0     & 1     & 3     & 0     & 3 \\
    2015/16 & 2     & 1     & 1     & 0     & 1     & $-$1    & 2     & 0     & 2     & 0     & 2     & $-$2    & 0     & 1     & $-$1    & 3     & 0     & 3 \\
    2016/17 & 2     & 1     & 1     & 1     & 1     & 0     & 1     & 2     & $-$1    & 2     & 0     & 2     & 0     & 2     & $-$2    & 2     & 2     & 0 \\
    2017/18 & 4     & 1     & 3     & 1     & 0     & 1     & 0     & 1     & $-$1    & 1     & 1     & 0     & 0     & 1     & $-$1    & 1     & 2     & $-$1 \\
    2018/19 & 1     & 3     & $-$2    & 1     & 1     & 0     & 2     & 1     & 1     & 1     & 1     & 0     & 1     & 0     & 1     & 2     & 1     & 1 \\
    2019/20 & 2     & 2     & 0     & 1     & 1     & 0     & 2     & 1     & 1     & 1     & 2     & $-$1    & 0     & 0     & 0     & 2     & 2     & 0 \\
    2020/21 & 3     & 0     & 3     & 1     & 0     & 1     & 2     & 2     & 0     & 1     & 2     & $-$1    & 0     & 1     & $-$1    & 1     & 3     & $-$2 \\
    2021/22 & 3     & 1     & 2     & 1     & 1     & 0     & 1     & 0     & 1     & 1     & 1     & 0     & 0     & 2     & $-$2    & 1     & 2     & $-$1 \\
    2022/23 & 3     & 1     & 2     & 0     & 1     & $-$1    & 1     & 3     & $-$2    & 1     & 2     & $-$1    & 2     & 0     & 2     & 1     & 0     & 1 \\
    2023/24 & 2     & 0     & 2     & 0     & 1     & $-$1    & 2     & 1     & 1     & 0     & 3     & $-$3    & 0     & 1     & $-$1    & 4     & 0     & 4 \\ \toprule
    Average & \multicolumn{3}{c}{1.19} & \multicolumn{3}{c}{0} & \multicolumn{3}{c}{0.05} & \multicolumn{3}{c}{$-$0.19} & \multicolumn{3}{c}{$-$0.38} & \multicolumn{3}{c}{0.67} \\ \bottomrule
	\end{tabularx}
\begin{tablenotes} \footnotesize
\item
\emph{Notes}: Column 1 shows the number of group winners; Column 2 shows the number of runners-up; Column $\Delta$ shows the difference between the number of group winners and runners-up.
\item
The last row shows the average of Column $\Delta$ for each country.
\item
The number of exclusions for any national association equals the number of group winners times the number of runners-up. The only exception is the 2005/06 season when Chelsea and Liverpool both qualified from Group G, thus, the number of constraints due to England is one rather than two. In addition, a Russian group winner could not have played against a Ukrainian runner-up in 2015/16.
\end{tablenotes}
\end{threeparttable}
}
\end{table}

However, this nice symmetry disappears in the UEFA Champions League, where the two sides are given by the group winners and the runner-up, and the type constraints are implied by the national associations of the clubs.
According to Table~\ref{Table4}, six leagues have had more than one club qualified for the Round of 16 in at least one year between 2003/04 and 2023/24. Due to England and, to a lesser extent, Spain, there are usually more group winners in these countries than runners-up. Consequently, the set $U$ is more likely to correspond to the runners-up when, on the basis of Figure~\ref{Fig9a}, the Standard Drop procedure is better, that is, the draw needs to start on the side of the runners-up.

\begin{result} \label{Result4}
For the graphs corresponding to the historical draws of the UEFA Champions League Round of 16, the official UEFA draw mechanism is expected to be fairer than its variant that draws the group winners first.
\end{result}

Result~\ref{Result4} can be illustrated by the finding that the cumulated $\mathit{MFD}$ ($\mathit{AFD}$) over the 21 seasons is smaller by 10.9\% (3.3\%) under the official mechanism than under its version where the group winners are drawn first.

According to Results~\ref{Result3} and \ref{Result4}, UEFA has chosen the optimal randomisation procedure among the four field-proven alternatives. This substantially refines the main result of \citet{BoczonWilson2023} about the near-optimality of the official UEFA draw mechanism, which is not only closely fair but also outperforms all reasonable field-proven variants. Naturally, previous studies \citep{BoczonWilson2023, KlossnerBecker2013} have not identified the advantage of the Drop mechanism over the Skip mechanism since not considering the latter, while both \citep[Footnote~33]{BoczonWilson2023} and \citet[Footnote~19]{KlossnerBecker2013} suggest that the difference between the Standard and Reversed Drop mechanisms is slight and uninteresting.

\subsection{Graphs with highly concentrated constraints} \label{Sec56}

The Reversed Drop mechanism is the most unfair among the historical Champions League seasons in 2017/18, when the difference between the Standard and Reversed forms is the highest, too. The associated bipartite graph is $I_1$ in Figure~\ref{Fig3}. We have created seven further graphs having a similar structure, which are also shown in Figure~\ref{Fig3}. All of them contain at least one node with four (except for $I_3$) or two nodes with three ($I_3$) type constraints.

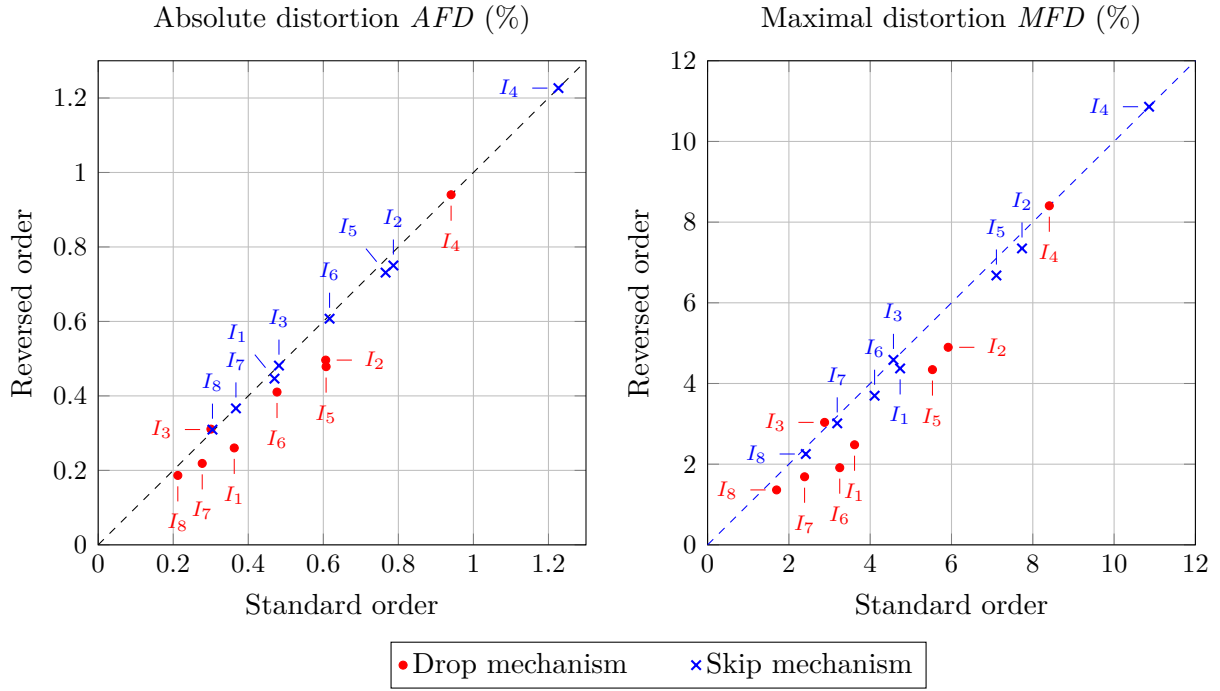
\begin{figure}[t!]
\centering

\begin{tikzpicture}
\begin{axis}[
name = axis1,
title = {Absolute distortion $\mathit{AFD}$ (\%)},
title style = {font=\small},
xlabel = $\mathit{AFD}$ of the Standard order,
x label style = {font=\small},
ylabel = $\mathit{AFD}$ of the Reversed order,
y label style = {font=\small},
width = 0.5\textwidth,
height = 0.5\textwidth,
nodes near coords,
xmajorgrids = true,
ymajorgrids = true,
xmin = 0,
xmax = 1.3,
ymin = 0,
ymax = 1.3,
]
\addplot [scatter,red,only marks,mark size=1.5pt,point meta=explicit symbolic] coordinates {
(0.362579669058374,0.260266248942994)
(0.606312825736648,0.496180950736648)
(0.300423537342155,0.310807620675489)
(0.94048946588799,0.940205721207138)
(0.607375613010739,0.478552130383974)
(0.476576911249559,0.410367422122957)
(0.277228684771103,0.218681711376466)
(0.21222286748029,0.186451868452316)
};
\draw [black,dashed] (rel axis cs:0,0) -- (rel axis cs:1,1);
\node[pin={[pin distance=0.2cm, ultra thick, pin edge={red}] 270:{\textcolor{red}{\scriptsize{$I_1$}}}}] at (0.362579669058374,0.260266248942994) {};
\node[pin={[pin distance=0.2cm, ultra thick, pin edge={red}] 360:{\textcolor{red}{\scriptsize{$I_2$}}}}] at (0.606312825736648,0.496180950736648) {};
\node[pin={[pin distance=0.2cm, ultra thick, pin edge={red}] 180:{\textcolor{red}{\scriptsize{$I_3$}}}}] at (0.300899370675489,0.309640912342156) {};
\node[pin={[pin distance=0.2cm, ultra thick, pin edge={red}] 270:{\textcolor{red}{\scriptsize{$I_4$}}}}] at (0.94048946588799,0.940205721207138) {};
\node[pin={[pin distance=0.2cm, ultra thick, pin edge={red}] 270:{\textcolor{red}{\scriptsize{$I_5$}}}}] at (0.607375613010738,0.478552130383974) {};
\node[pin={[pin distance=0.2cm, ultra thick, pin edge={red}] 270:{\textcolor{red}{\scriptsize{$I_6$}}}}] at (0.476576911249559,0.410367422122957) {};
\node[pin={[pin distance=0.2cm, ultra thick, pin edge={red}] 270:{\textcolor{red}{\scriptsize{$I_7$}}}}] at (0.277228684771103,0.218681711376466) {};
\node[pin={[pin distance=0.2cm, ultra thick, pin edge={red}] 270:{\textcolor{red}{\scriptsize{$I_8$}}}}] at (0.21222286748029,0.186451868452316) {};
\addplot [scatter,blue,only marks,mark=x,mark size=2.5pt,thick,point meta=explicit symbolic] coordinates {
(0.469816248942994,0.446424248942994)
(0.786677617403314,0.750255659069981)
(0.481761079008822,0.481402802326587)
(1.2265759339731,1.22658078503693)
(0.765961789958443,0.731111490246786)
(0.616646911249559,0.607336145292112)
(0.367082378043133,0.366492794709799)
(0.304516555783561,0.309439533210929)
};
\node[pin={[pin distance=0.2cm, ultra thick, pin edge={blue}] 120:{\textcolor{blue}{\scriptsize{$I_1$}}}}] at (0.469816248942994,0.446424248942994) {};
\node[pin={[pin distance=0.2cm, ultra thick, pin edge={blue}] 90:{\textcolor{blue}{\scriptsize{$I_2$}}}}] at (0.786677617403314,0.750255659069981) {};
\node[pin={[pin distance=0.2cm, ultra thick, pin edge={blue}] 90:{\textcolor{blue}{\scriptsize{$I_3$}}}}] at (0.481761079008822,0.481402802326587) {};
\node[pin={[pin distance=0.2cm, ultra thick, pin edge={blue}] 180:{\textcolor{blue}{\scriptsize{$I_4$}}}}] at (1.2265759339731,1.22658078503693) {};
\node[pin={[pin distance=0.2cm, ultra thick, pin edge={blue}] 120:{\textcolor{blue}{\scriptsize{$I_5$}}}}] at (0.765961789958443,0.731111490246786) {};
\node[pin={[pin distance=0.2cm, ultra thick, pin edge={blue}] 90:{\textcolor{blue}{\scriptsize{$I_6$}}}}] at (0.616646911249559,0.607336145292112) {};
\node[pin={[pin distance=0.2cm, ultra thick, pin edge={blue}] 90:{\textcolor{blue}{\scriptsize{$I_7$}}}}] at (0.367082378043133,0.366492794709799) {};
\node[pin={[pin distance=0.2cm, ultra thick, pin edge={blue}] 90:{\textcolor{blue}{\scriptsize{$I_8$}}}}] at (0.304516555783561,0.309439533210929) {};
\end{axis}

\begin{axis}[
at = {(axis1.south east)},
xshift = 0.1\textwidth,
title = {Maximal distortion $\mathit{MFD}$ (\%)},
title style = {font=\small},
xlabel = $\mathit{MFD}$ of the Standard order,
x label style = {font=\small},
ylabel = $\mathit{MFD}$ of the Reversed order,
y label style = {font=\small},
width = 0.5\textwidth,
height = 0.5\textwidth,
nodes near coords,
xmajorgrids = true,
ymajorgrids = true,
xmin = 0,
xmax = 12,
ymin = 0,
ymax = 12,
legend style = {font=\small,at={(-0.65,-0.2)},anchor=north west,legend columns=3},
legend entries = {Drop mechanism$\qquad$,Skip mechanism}
]
\addplot [scatter,red,only marks,mark size=1.5pt,point meta=explicit symbolic] coordinates {
(3.61513049551676,2.48108549551675)
(5.91848361325967,4.89715361325967)
(2.87850603113648,3.03629703113648)
(8.40618744836776,8.40517144836775)
(5.53118871888203,4.34370771888203)
(3.25062201104973,1.91257101104972)
(2.38603470150833,1.68764170150834)
(1.69718396954315,1.36173696954315)
};
\draw [blue,dashed] (rel axis cs:0,0) -- (rel axis cs:1,1);
\node[pin={[pin distance=0.2cm, ultra thick, pin edge={red}] 270:{\textcolor{red}{\scriptsize{$I_1$}}}}] at (3.61513049551676,2.48108549551675) {};
\node[pin={[pin distance=0.2cm, ultra thick, pin edge={red}] 360:{\textcolor{red}{\scriptsize{$I_2$}}}}] at (5.91848361325967,4.89715361325967) {};
\node[pin={[pin distance=0.2cm, ultra thick, pin edge={red}] 180:{\textcolor{red}{\scriptsize{$I_3$}}}}] at (2.87850603113648,3.03629703113648) {};
\node[pin={[pin distance=0.2cm, ultra thick, pin edge={red}] 270:{\textcolor{red}{\scriptsize{$I_4$}}}}] at (8.40618744836776,8.40517144836775) {};
\node[pin={[pin distance=0.2cm, ultra thick, pin edge={red}] 270:{\textcolor{red}{\scriptsize{$I_5$}}}}] at (5.53118871888203,4.34370771888203) {};
\node[pin={[pin distance=0.2cm, ultra thick, pin edge={red}] 270:{\textcolor{red}{\scriptsize{$I_6$}}}}] at (3.25062201104973,1.91257101104972) {};
\node[pin={[pin distance=0.2cm, ultra thick, pin edge={red}] 270:{\textcolor{red}{\scriptsize{$I_7$}}}}] at (2.38603470150833,1.68764170150834) {};
\node[pin={[pin distance=0.2cm, ultra thick, pin edge={red}] 180:{\textcolor{red}{\scriptsize{$I_8$}}}}] at (1.69718396954315,1.36173696954315) {};
\addplot [scatter,blue,only marks,mark=x,mark size=2.5pt,thick,point meta=explicit symbolic] coordinates {
(4.73620549551675,4.37280249551675)
(7.73453161325967,7.34826061325967)
(4.57178503113648,4.58633003113648)
(10.8605964483678,10.8605784483678)
(7.10442771888203,6.67784971888203)
(4.10838101104972,3.69726501104972)
(3.18875970150833,3.01230370150833)
(2.41598696954314,2.25215396954315)
};
\node[pin={[pin distance=0.2cm, ultra thick, pin edge={blue}] 270:{\textcolor{blue}{\scriptsize{$I_1$}}}}] at (4.73620549551675,4.37280249551675) {};
\node[pin={[pin distance=0.2cm, ultra thick, pin edge={blue}] 90:{\textcolor{blue}{\scriptsize{$I_2$}}}}] at (7.73453161325967,7.34826061325967) {};
\node[pin={[pin distance=0.2cm, ultra thick, pin edge={blue}] 90:{\textcolor{blue}{\scriptsize{$I_3$}}}}] at (4.57178503113648,4.58633003113648) {};
\node[pin={[pin distance=0.2cm, ultra thick, pin edge={blue}] 180:{\textcolor{blue}{\scriptsize{$I_4$}}}}] at (10.8605964483678,10.8605784483678) {};
\node[pin={[pin distance=0.2cm, ultra thick, pin edge={blue}] 90:{\textcolor{blue}{\scriptsize{$I_5$}}}}] at (7.10442771888203,6.67784971888203) {};
\node[pin={[pin distance=0.2cm, ultra thick, pin edge={blue}] 90:{\textcolor{blue}{\scriptsize{$I_6$}}}}] at (4.10838101104972,3.69726501104972) {};
\node[pin={[pin distance=0.2cm, ultra thick, pin edge={blue}] 90:{\textcolor{blue}{\scriptsize{$I_7$}}}}] at (3.18875970150833,3.01230370150833) {};
\node[pin={[pin distance=0.2cm, ultra thick, pin edge={blue}] 180:{\textcolor{blue}{\scriptsize{$I_8$}}}}] at (2.41598696954314,2.25215396954315) {};
\end{axis}
\end{tikzpicture}

\caption{Fairness distortions of the draw mechanisms for the graphs in Figure~\ref{Fig3}}
\label{Fig10}

\end{figure}


As can be seen in Figure~\ref{Fig10}, the draw mechanisms are relatively unfair for these instances. The maximal error is above 4 for three graphs, $I_2$, $I_4$, and $I_5$.

\begin{table}[t!]
  \centering
  \caption{Fairness distortions for the graph $I_4$ in Figure~\ref{Fig3}}
  \label{Table5}
  
\begin{subtable}{\textwidth}
\centering
  \caption{Standard and Reversed Drop mechanisms}
  \label{Table5a}
\begin{footnotesize}
    \begin{tabularx}{0.8\textwidth}{l CCCC CCCC} \toprule
          & V1    & V2    & V3    & V4    & V5    & V6    & V7    & V8 \\    \bottomrule
    U1    & X     & \cellcolor{ForestGreen!2.41819764823453} 0.27 & \cellcolor{ForestGreen!2.37259914823453} 0.26 & \cellcolor{ForestGreen!2.36791914823452} 0.26 & \cellcolor{ForestGreen!2.3582171482345} 0.26 & X     & \cellcolor{red!4.75204729646902} 0.53 & \cellcolor{red!4.76488579646901} 0.53 \\
    U2    & X     & X     & \cellcolor{ForestGreen!2.37208164823453} 0.26 & \cellcolor{ForestGreen!2.36671764823453} 0.26 & \cellcolor{ForestGreen!2.41885014823451} 0.27 & \cellcolor{ForestGreen!2.37818814823453} 0.26 & \cellcolor{red!4.75286629646899} 0.53 & \cellcolor{red!4.78297129646901} 0.53 \\
    U3    & X     & \cellcolor{ForestGreen!2.39706114823452} 0.27 & X     & \cellcolor{ForestGreen!2.37040764823451} 0.26 & \cellcolor{ForestGreen!2.37259914823453} 0.26 & \cellcolor{ForestGreen!2.34314664823452} 0.26 & \cellcolor{red!4.76246929646901} 0.53 & \cellcolor{red!4.72074529646901} 0.52 \\
    U4    & X     & \cellcolor{ForestGreen!2.36875164823453} 0.26 & \cellcolor{ForestGreen!2.36229864823453} 0.26 & X     & \cellcolor{ForestGreen!2.35149414823452} 0.26 & \cellcolor{ForestGreen!2.41402164823453} 0.27 & \cellcolor{red!4.75197979646901} 0.53 & \cellcolor{red!4.744586296469} 0.53 \\
    U5    & X     & \cellcolor{ForestGreen!2.36296464823453} 0.26 & \cellcolor{ForestGreen!2.41033164823453} 0.27 & \cellcolor{ForestGreen!2.39874414823451} 0.27 & X     & \cellcolor{ForestGreen!2.37669414823452} 0.26 & \cellcolor{red!4.75454479646903} 0.53 & \cellcolor{red!4.794189796469} 0.53 \\
    U6    & \cellcolor{red!75.6511150353099} 8.41 & X     & X     & X     & X     & X     & \cellcolor{ForestGreen!37.8382790176549} 4.2 & \cellcolor{ForestGreen!37.8128360176549} 4.2 \\
    U7    & \cellcolor{ForestGreen!37.7889500176549} 4.2 & \cellcolor{red!4.74826729646901} 0.53 & \cellcolor{red!4.77108229646899} 0.53 & \cellcolor{red!4.75655179646902} 0.53 & \cellcolor{red!4.74152629646902} 0.53 & \cellcolor{red!4.76606479646901} 0.53 & X     & \cellcolor{red!14.0054575353098} 1.56 \\
    U8    & \cellcolor{ForestGreen!37.8621650176549} 4.21 & \cellcolor{red!4.79870779646902} 0.53 & \cellcolor{red!4.74622879646901} 0.53 & \cellcolor{red!4.74723679646901} 0.53 & \cellcolor{red!4.75963429646903} 0.53 & \cellcolor{red!4.74598579646899} 0.53 & \cellcolor{red!14.0643715353098} 1.56 & X \\ \toprule
    \end{tabularx}
\end{footnotesize}
\end{subtable}

\vspace{0.25cm}
\begin{subtable}{\textwidth}
\centering
  \caption{Standard and Reversed Skip mechanisms}
  \label{Table5b}
\begin{threeparttable}
\begin{footnotesize}
    \begin{tabularx}{0.8\textwidth}{l CCCC CCCC} \toprule
          & V1    & V2    & V3    & V4    & V5    & V6    & V7    & V8 \\    \bottomrule
    U1    & X     & \cellcolor{ForestGreen!3.22351314823454} 0.36 & \cellcolor{ForestGreen!3.18642864823452} 0.35 & \cellcolor{ForestGreen!3.22723464823453} 0.36 & \cellcolor{ForestGreen!3.18784614823452} 0.35 & X     & \cellcolor{red!6.41405479646902} 0.71 & \cellcolor{red!6.41096779646902} 0.71 \\
    U2    & X     & X     & \cellcolor{ForestGreen!3.17782914823453} 0.35 & \cellcolor{ForestGreen!3.1761281482345} 0.35 & \cellcolor{ForestGreen!3.20212014823451} 0.36 & \cellcolor{ForestGreen!3.19995564823452} 0.36 & \cellcolor{red!6.36582829646902} 0.71 & \cellcolor{red!6.390204796469} 0.71 \\
    U3    & X     & \cellcolor{ForestGreen!3.20128314823452} 0.36 & X     & \cellcolor{ForestGreen!3.15982464823451} 0.35 & \cellcolor{ForestGreen!3.20464014823453} 0.36 & \cellcolor{ForestGreen!3.19214364823454} 0.35 & \cellcolor{red!6.37196179646901} 0.71 & \cellcolor{red!6.38592979646901} 0.71 \\
    U4    & X     & \cellcolor{ForestGreen!3.16089114823452} 0.35 & \cellcolor{ForestGreen!3.17257314823452} 0.35 & X     & \cellcolor{ForestGreen!3.20319114823454} 0.36 & \cellcolor{ForestGreen!3.24866814823453} 0.36 & \cellcolor{red!6.40825429646902} 0.71 & \cellcolor{red!6.377069296469} 0.71 \\
    U5    & X     & \cellcolor{ForestGreen!3.19733214823453} 0.36 & \cellcolor{ForestGreen!3.24415914823451} 0.36 & \cellcolor{ForestGreen!3.19961364823453} 0.36 & X     & \cellcolor{ForestGreen!3.16537314823451} 0.35 & \cellcolor{red!6.43839979646902} 0.72 & \cellcolor{red!6.36807829646903} 0.71 \\
    U6    & \cellcolor{red!97.7452870353098} 10.86 & X     & X     & X     & X     & X     & \cellcolor{ForestGreen!48.8971325176549} 5.43 & \cellcolor{ForestGreen!48.8481545176549} 5.43 \\
    U7    & \cellcolor{ForestGreen!48.8452340176549} 5.43 & \cellcolor{red!6.38790529646902} 0.71 & \cellcolor{red!6.37655629646902} 0.71 & \cellcolor{red!6.37670479646902} 0.71 & \cellcolor{red!6.37152979646902} 0.71 & \cellcolor{red!6.41663329646901} 0.71 & X     & \cellcolor{red!16.9159045353098} 1.88 \\
    U8    & \cellcolor{ForestGreen!48.9000530176549} 5.43 & \cellcolor{red!6.39511429646899} 0.71 & \cellcolor{red!6.40443379646901} 0.71 & \cellcolor{red!6.38609629646902} 0.71 & \cellcolor{red!6.42626779646901} 0.71 & \cellcolor{red!6.38950729646902} 0.71 & \cellcolor{red!16.8986335353098} 1.88 & X \\ \toprule
    \end{tabularx}
\end{footnotesize}
    \begin{tablenotes} \footnotesize
\item
X represents a pair of teams that cannot play in the Round of 16.
\item
The numbers show percentages ($100 \cdot \left( p_{ij} - p_{ij}^M \right)$) rounded to two decimal places.
\item
\textcolor{ForestGreen}{Green} (\textcolor{red}{Red}) colour means that the draw mechanism implies a higher (lower) probability than the Uniform mechanism.
\item
Darker colour indicates a higher value.
    \end{tablenotes}
\end{threeparttable}
\end{subtable}
\end{table}


Table~\ref{Table5} presents the distortions for graph $I_4$, which coincide for the Standard and Reversed variants of the two main mechanisms since the sets $U$ and $V$ are exchangeable.
The implied $\mathit{MFD}$ is worrying because the nodes with five restrictions have three possible nodes to be assigned to them, thus, every reasonable assignment rule should match these two nodes with at least a probability of $1/3$. Consequently, a na\"ive upper bound of maximal fairness distortion is $61.76 - 33.33 \approx 28.4$(\%), and the Drop mechanism is unable to reduce this value by more than 71\%. Furthermore, the value of $\mathit{MFD}$ exceeds the worst maximal fairness distortion for all graphs considered in Section~\ref{Sec52} (see Figure~\ref{Fig5}), that is, the Round of 16 draw can be less fair than a potential draw of quarterfinals under the same rules.

\begin{result} \label{Result5}
The Drop mechanism does not come quantitatively close to the best possible lottery for certain bipartite graphs, which represent possible UEFA Champions League seasons and have some nodes with a high number of type constraints on both sides.
\end{result}

Result~\ref{Result5} somewhat contradicts \citet[Result~3]{BoczonWilson2023} that the Drop procedure continues to be close to a constrained-best if (i) the number of constraints; (ii) the likely location of the constraints in the graph; and (iii) the number of nodes in the graph is shifted. The explanation is straightforward because randomly creating the set of exclusions will result in a graph similar to $I_2$, $I_4$, or $I_5$ with only a small probability, and averaging the biases over hundreds of such instances covers the high distortions for some particular graphs.

Figure~\ref{Fig10} reinforces the message of Section~\ref{Sec55}.
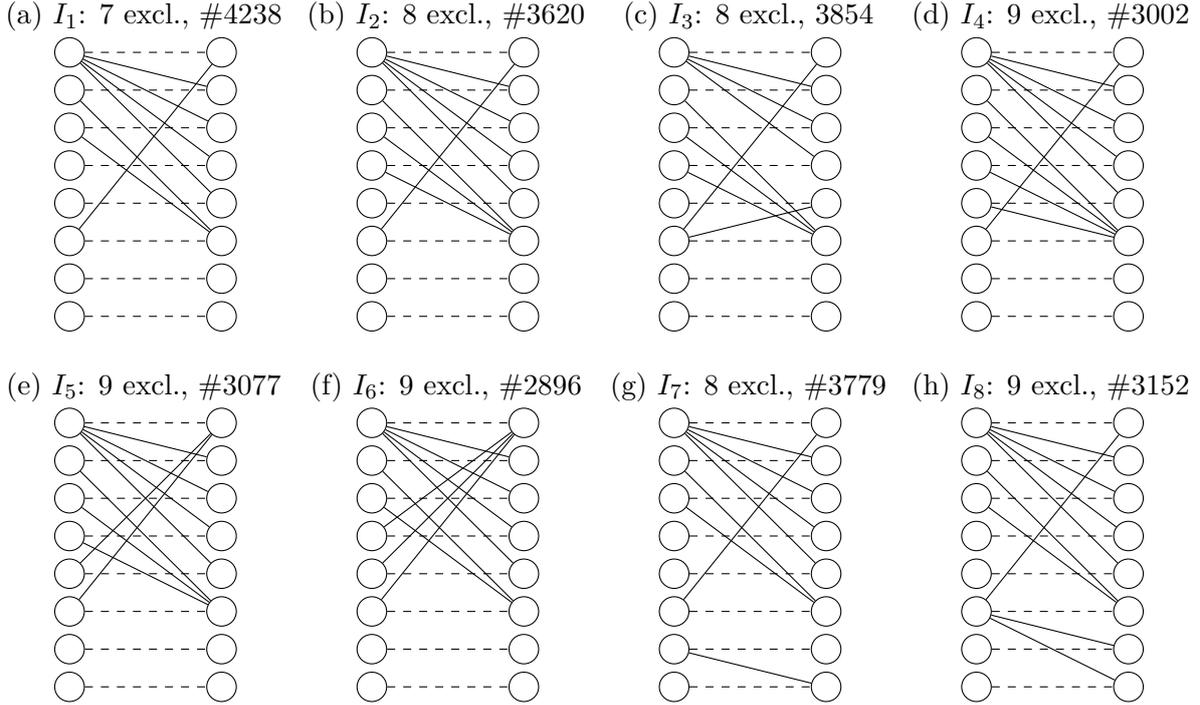
First, the Drop mechanism is clearly fairer compared to the Skip mechanism, it has a lower absolute fairness distortion by about 30--40\% for our balanced bipartite graphs with 16 nodes. In addition, the draw order has a more moderated impact on the Skip mechanism.
Second, the two versions of the Drop mechanism can substantially differ from each other, the distortion of the Reversed Drop mechanism is smaller by at least 17\% for the three ``unfavourable'' graphs of $I_1$, $I_2$, and $I_5$. Furthermore, the $\mathit{MFD}$ of the Standard Drop mechanism is above 3.2(\%) for graph $I_6$ but remains below 2(\%) under the Reversed Drop, implying a decrease of over 40\%.

\begin{result} \label{Result6}
The draw order of the Drop mechanism sometimes has a non-negligible effect on the level of unfairness.
\end{result}

According to \citet[Footnote~19]{KlossnerBecker2013}, the probabilities of the UEFA procedure would change \emph{slightly} if group winners were drawn first and only then matched with suitable runners-up. Result~\ref{Result6} shows that it is a rough conclusion and may be true only for the Skip mechanism.

Last but not least, Tables~\ref{Table3} and \ref{Table5} imply suggest a general pattern that is reinforced by a thorough study of distortions.

\begin{result} \label{Result7}
The four draw mechanisms (Standard/Reversed Drop, Standard/Reversed Skip) do no fundamentally differ in the sense that they usually change the probabilities under the Uniform draw in the same direction and roughly with the same magnitude.
\end{result}

It requires further research whether Result~\ref{Result7} holds for other randomisation procedures or not.

\section{Conclusions} \label{Sec6}

Our paper has compared the basic randomisation procedures used in sports competitions to divide some teams into groups in the presence of draw constraints. Since uniform distribution over all valid assignments should be sacrificed for the sake of transparency, it is crucial to know which variant (Standard or Reversed) of what mechanism (Drop or Skip) is the least distorted in a given setting. Therefore, we have evaluated their unfairness for selected sets of balanced bipartite graphs with two kinds of restrictions. According to our results, the UEFA Champions League Round of 16 draw has adopted the optimal transparent field-proven algorithm. On the other hand,
(1) this is not the optimal choice in every setting, for instance, for a draw of quarterfinals under the same rules; and
(2) there remains a non-negligible scope to find fairer randomisation procedures, especially since they tend to distort the probabilities in the same direction and roughly with the same magnitude.

There are several promising directions for future research. First, the reason for the difference between the Drop and Skip mechanisms has not been fully explored, although the difficulty of the problem has been highlighted. Second, these draw procedures are worth analysing in further cases. The advantage of the Drop mechanism in bipartite graphs suggests that it might be a competitive option for the FIFA World Cup draw---however, having groups with more than two teams increases the complexity of the potential constraints.
Last but not least, in contrast to \citet{BoczonWilson2023}, we think tournament organisers should continue the quest for fairer but transparent lotteries since all previous proposals \citep{Guyon2014a, KlossnerBecker2013, LalienaLopez2025, RobertsRosenthal2024} have different weaknesses.


\section*{Acknowledgements}
\addcontentsline{toc}{section}{Acknowledgements}
\noindent
This paper could not have been written without my \emph{father} (also called \emph{L\'aszl\'o Csat\'o}), who has helped to code the simulations in Python. \\
Ten anonymous colleagues and \emph{Ilia Tsetlin} have provided valuable comments and suggestions on earlier drafts. \\
The research was supported by the National Research, Development and Innovation Office under Grant FK 145838 and the J\'anos Bolyai Research Scholarship of the Hungarian Academy of Sciences.

\bibliographystyle{apalike}
\bibliography{All_references}

\end{document}